\documentclass[12pt,a4paper,english]{article}
\usepackage[T1]{fontenc}
\usepackage[latin9]{inputenc}
\usepackage{babel}
\usepackage{amsmath}
\usepackage{amssymb}
\usepackage[unicode=true,
 bookmarks=true,bookmarksnumbered=false,bookmarksopen=false,
 breaklinks=false,pdfborder={0 0 1},colorlinks=false]
 {hyperref}
 \usepackage{color}
\usepackage{breakurl}

\makeatletter

\providecommand{\tabularnewline}{\\}
\usepackage{a4}\usepackage[all]{xy}
\usepackage{multirow}
\usepackage{cite}

\usepackage{babel}
\usepackage{youngtab}

\makeatother

\begin{document}
\vspace*{-1.5cm}

\global\long\def\todo#1{{\bf ?????!!!! #1 ?????!!!!}\marginpar{$\Longleftarrow$}}
 \global\long\def\fref#1{Figure~\ref{#1}}
 \global\long\def\tref#1{Table~\ref{#1}}
 \global\long\def\sref#1{\S~\ref{#1}}
 \global\long\def\nn{\nonumber}
 \global\long\def\tr{\mathop{\rm Tr}}
 \global\long\def\comment#1{}

\global\long\def\cM{{\cal M}}
 \global\long\def\cW{{\cal W}}
 \global\long\def\cN{{\cal N}}
 \global\long\def\cH{{\cal H}}
 \global\long\def\cK{{\cal K}}
 \global\long\def\cZ{{\cal Z}}
 \global\long\def\cO{{\cal O}}
 \global\long\def\cB{{\cal B}}
 \global\long\def\cC{{\cal C}}
 \global\long\def\cD{{\cal D}}
 \global\long\def\cE{{\cal E}}
 \global\long\def\cF{{\cal F}}
 \global\long\def\cX{{\cal X}}
 \global\long\def\IA{\mathbb{A}}
 \global\long\def\IP{\mathbb{P}}
 \global\long\def\IQ{\mathbb{Q}}
 \global\long\def\IH{\mathbb{H}}
 \global\long\def\IR{\mathbb{R}}
 \global\long\def\IC{\mathbb{C}}
 \global\long\def\IF{\mathbb{F}}
 \global\long\def\IV{\mathbb{V}}
 \global\long\def\II{\mathbb{I}}
 \global\long\def\IZ{\mathbb{Z}}
 \global\long\def\re{{\rm Re}}
 \global\long\def\im{{\rm Im}}
 \global\long\def\li{{\rm Li}}

\global\long\def\CA{\mathbb{A}}
 \global\long\def\CP{\mathbb{P}}
 \global\long\def\tmat#1{{\tiny\left(\begin{matrix} #1 \end{matrix}\right)}}
 \global\long\def\mat#1{\left(\begin{matrix} #1 \end{matrix}\right)}

\global\long\def\diff#1#2{\frac{\partial#1}{\partial#2}}
 \global\long\def\gen#1{\langle#1 \rangle}

\global\long\def\drawsquare#1#2{\hbox{%
\rule{#2pt}{#1pt}\hskip-#2pt
\rule{#1pt}{#2pt}\hskip-#1pt
\rule[#1pt]{#1pt}{#2pt}}\rule[#1pt]{#2pt}{#2pt}\hskip-#2pt
\rule{#2pt}{#1pt}}
\global\long\def\fund{\raisebox{-.5pt}{\drawsquare{6.5}{0.4}}}
 \global\long\def\antifund{\overline{\fund}}

\newtheorem{theorem}{\textbf{THEOREM} }\global\long\def\thetheorem{\thesection.\arabic{theorem}}
 \textbf{ \newtheorem{proposition}{PROPOSITION} }\global\long\def\thetheorem{\thesection.\arabic{proposition}}
 \textbf{ \newtheorem{observation}{OBSERVATION} }\global\long\def\thetheorem{\thesection.\arabic{observation}}

\global\long\def\theequation{\thesection.\arabic{equation}}
 \textbf{ }\global\long\def\setall{\setcounter{equation}{0} \setcounter{theorem}{0}}
 \textbf{ }\global\long\def\setequation{\setcounter{equation}{0}}

\vspace{1cm}

\comment{
\begin{center}
Numerical Algebraic Geometry: A New Perspective on Gauge and String theories
\par\end{center}
\vspace{0.75cm}
\begin{center}
Dhagash Mehta
\footnote{Email: dbmehta@syr.edu}
\par\end{center}
\vspace{0.15cm}
\begin{center}
Physics Department, Syracuse University, Syracuse, NY 13244, USA.
\par\end{center}
}

~\\
\vskip 1cm

\centerline{{\Large \bf Numerical Algebraic Geometry: A New}}
~\\
\centerline{{\Large \bf Perspective on String and Gauge Theories}}
\medskip

\vspace{.4cm}

\centerline{
{\large Dhagash Mehta,}$^1$
{\large Yang-Hui He}$^2$ \&
{\large Jonathan D. Hauenstein}$^3$
}
\vspace*{2.0ex}

\begin{center}
{\it
{\small
{${}^{1}$
Department of Physics, Syracuse University, Syracuse, NY 13244, USA.\\
email: dbmehta@syr.edu \\
}
\vspace*{1.5ex}
{${}^{2}$ Department of Mathematics, City University, London, EC1V 0HB, UK;\\
School of Physics, NanKai University, Tianjin, 300071, P.R.~China; \\
Merton College, University of Oxford, OX14JD, UK\\
email: hey@maths.ox.ac.uk \\
}
\vspace*{1.5ex}
{${}^{3}$ Department of Mathematics, Texas A\&M University, College Station, TX 77843-3368, USA. \\
email: jhauenst@math.tamu.edu \\
}
}}
\end{center}

\begin{abstract}
\noindent
The interplay rich between algebraic geometry and string and gauge theories has recently been immensely aided by advances
in computational algebra.
However, these symbolic (Gr\"obner) methods are severely limited by algorithmic issues such as exponential space complexity
and being highly sequential. In this paper, we introduce a novel paradigm of numerical algebraic geometry which in a plethora of situations overcomes these shortcomings. Its so-called 'embarrassing parallelizability' allows us to solve many problems and extract physical information which elude the symbolic methods. We describe the method and then use it to solve various problems arising from physics which could not be otherwise solved.
\end{abstract}


\newpage

 \renewcommand{\baselinestretch}{1.5}
\tableofcontents
 \renewcommand{\baselinestretch}{1}

\newpage

\section{Introduction}
Increasingly is the field of computational algebraic geometry becoming
a significant tool to contemporary theoretical physics \cite{issue}.
This is becoming ever more transparent especially in string theory and in gauge theory;
indeed, the plethora of vacuum configurations in string and M-theory is clearly best suited
for large-scale intensive computerization \cite{grant},
so too can new hidden structures in gauge theory be
uncovered by computing new geometric invariants (cf.~e.g.\cite{Gray:2005sr,Anderson:2007nc,Gray:2007yq}).

The rapid progress in algorithmic geometry as well as computer power has been a
crucial ingredient to such advances.
Key to much of the algorithms analyzing simultaneous multi-variate polynomial systems of high
degree, to which the aforementioned physical situations can ultimately be reduced, is
the Gr\"obner basis technique (for an excellent review in this context, cf.~\cite{Gray:2009fy}).
Roughly speaking, given the ideal generated by the vanishing set of the polynomials,
the so-called Buchberger Algorithm (BA) or its refined variants
can compute a new system of equations, called a Gr\"obner basis \cite{CLO:07}.

The BA reduces to Gaussian elimination in
the case of linear equations.
Similarly it is also a generalization of the Euclidean algorithm
for the computation of the Greatest Common Divisors of a univariate
polynomial. Recently, more efficient variants of the BA have been
developed to obtain a Gr\"obner basis, e.g., F4 \cite{Faugere99anew},
F5 \cite{Faug:02} and Involution Algorithms~\cite{2005math1111G}.
Symbolic computation packages such as {\sf Mathematica}, {\sf Maple}, {\sf Reduce},
etc., have built-in commands to calculate a Gr\"obner basis. Moreover, {\sf Singular}
\cite{DGPS}, {\sf COCOA} \cite{CocoaSystem} and {\sf Macaulay2} \cite{M2}
are specialized packages for Gr\"obner basis and Computational Algebraic
Geometry, available as freeware, and {\sf MAGMA} \cite{BCP:97} is also such
a specialized package available commercially.

In \cite{Gray:2006gn,Gray:2008zs,Gray:2007yq},
the power of Gr\"obner basis methods was harnessed for the sake of answering
questions pertinent to string and particle phenomenology.
A publicly available computational package, called {\sf Stringvacua},
is designed to interface {\sf Mathematica} with {\sf Singular} and has built-in utilities for string phenomenology.
By extracting important information such as the dimension of the ideal, the number of real roots
in the system, etc., many issues such as stability or supersymmetry of the potential or the branches of
moduli space of vacua, can be settled using only a regular desktop machine.

However, even with such reassuring techniques, there are a few problems with
symbolic methods: the BA is known to suffer from {\it exponential complexity}, both the RAM required by the
machine and the computation time increases exponentially with the number of variables, equations, degree and terms in each polynomial.
Thus, often for even seemingly small sized systems, one
may not be able to compute a Gr\"obner basis.
Furthermore, BA is also usually less
efficient for systems with irrational coefficients and one habitually has to resort to randomizing over the
space of integer (or prime) coefficients and work over finite fields.
Another drawback is the highly sequential nature of the BA and thus efficient
parallelizations currently do not exist.

When the information we are after is not sophisticated quantities such as cohomology of
vector bundles, which, for example is central to understanding particle spectra and Yukawa couplings
in heterotic compactifications, it is natural to wonder whether one could appeal to numerics.
Indeed, in a typical situation when one is confronted with the minimization of a potential to see
how many isolated solutions there are, what are their approximate values and whether supersymmetry
is preserved, numerical recipes specifically tailored to polynomial systems would seem ideal.

The purpose of this paper is to explain precisely such a method, called the {\bf Numerical Algebraic Geometry} (NAG) and
to introduce it in our context. The method was first used in lattice
field theory \cite{Mehta:2009} where
all the stationary points of a multivariate function called the lattice
Landau gauge fixing functional \cite{Mehta:2009zv,vonSmekal:2007ns,vonSmekal:2008es,Mehta:2010pe}
needed to be found.  We will see how the shortcomings of the Gr\"obner basis methods may be
overcome with this numerical method.

In general, the problems we encounter fall into two categories: (1) there is a finite number
of isolated extrema and (2) there is a continuum of solutions.
The first constitutes zero-dimensional ideals and arises, for example, when finding the number and
numerical positions of minima of a given potential, be it from a heterotic string model or from
a cosmological background.
The second gives us an affine algebraic variety and involves queries from an algebraic-geometric perspective.
These could come from analysis of the space of vacua of scalar fields in a supersymmetric
gauge theory, or from complex geometries arising from AdS/CFT.
The central method in NAG is the numerical polynomial homotopy continuation (NPHC):
it can be shown that for a given system of polynomial equations to be solved,
a homotopy between the system and a new system (which is easier to solve and share many
features with the former system) can be constructed (see \cite{Mehta:2011xs, Mehta:2011wj}
for a detailed description of this method in this context). Then, one tracks paths starting from
each solution of the new system as one moves towards the original system along the homotopy,
to finally obtain all the solutions of the original system.  The NPHC method is used to find all
the complex (which obviously include real) solutions of a system which is known to have only isolated solutions,
i.e., the first category of the systems. The NPHC method has been immensely useful in various areas of theoretical physics including statistical mechanics, particle phenomenology, string phenomenology, lattice field theories, etc. \cite{Mehta:2009,Mehta:2009zv,Mehta:2011xs,Mehta:2011wj,Kastner:2011zz,Kastner:11,Maniatis:2012ex,Casetti-Mehta:2012,Hughes-Mehta:2012}. For the second category of systems, the NPHC method can be
extended to describe the solution space completely (i.e., dimension, degree, and number of irreducible components).

Indeed, one of the greatest computational hurdles to Gr\"obner techniques
is the primary and irreducible decomposition of ideals which gives
crucial data of the branches of the moduli space.
We will address ample examples from both categories, demonstrating how many questions can
be settled by NAG.

The organization of the paper is as follows.
We begin in \sref{s:nphc} by describing the NPHC method, first pedagogically for a univariate example
before generalizing it to the multivariate polynomial ideal.
We then apply the method to the two categories of circumstances.
In \sref{s:0dim}, we treat several systems from heterotic and M-theory models which
the symbolic methods are  already known to be prohibitively difficult.
Next, in \sref{s:nag}, we discuss how to perform numerical irreducible decomposition to systems
which are positive dimensional, and apply to four different situations arising from string
and gauge theories. We conclude with prospects in \sref{s:conc}.

\section{The Numerical Polynomial Homotopy Continuation Method}\label{s:nphc}
In this section, we will outline one of the most powerful tools in numerical
algebraic geometry, the so-called {\em numerical polynomial homotopy continuation
method}, or NPHC.
Since this method is already introduced in particle physics and statistical mechanics areas in \cite{Mehta:2009,Mehta:2009zv,Mehta:2011xs,Mehta:2011wj,Kastner:2011zz,Kastner:11}, we will only give a brief explanation as to the principles and algorithms involved as well as
advertise some available softwares. We note that in these references only one kind of homotopy, called the Total Degree Homotopy,
was used whereas in the present paper we introduce two new kinds of homotopies, the 2-Homogeneous Homotopy and the Polyhedral Homotopy,
which in many ways are more powerful than the former.

We know from the Fundamental Theorem of Algebra that for a single variable polynomial equation of degree $k \in \IZ_{>0}$,
say $f(x)=\sum_{i=0}^{k}a_{i}x^{i}$, with coefficients $a_{i}$ and the variable $x$ both defined over $\mathbb{C}$,
the number of solutions is exactly $k$ if $a_{k}\neq0$, counting multiplicities.
To solve a univariate equation, many numerical methods such as the \textit{companion matrix}
trick for low degree polynomials, and divide-and-conquer techniques
for high degree polynomials, are available. Here, we introduce the NPHC method, which can then
be extended to the multivariate case in a straightforward manner \cite{SW:95,Li:2003}.

Briefly, in the NPHC method, first one writes down
the equation or system of equations to be solved in a more general
parametric form, then solves this system at a point in parameter space where
its solutions can be easily found, before finally tracking these solutions
from this point in parameter space to the point in parameter space
corresponding to the original system/problem. This approach can be
applied even to non-algebraic equations
which exhibit a continuous dependence of the solutions on the parameters. However,
there exist many difficulties, which do not arise in the polynomial setting,
in making this method a primary candidate method to solve a set of
non-algebraic equations.

Let us consider the univariate equation $z^{2}-5=0$, pretending that we do not know its solutions, i.e.,
$z=\pm\sqrt{5}$. We thus begin by defining the more general parametric family
\begin{equation}
H(z,t) = (1-t)(z^2 - 5) + \gamma t (z^2 - 1) = 0, \label{eq:one_var_homotopy}
\end{equation}
where $t\in[0,1]$ is a parameter and $\gamma$ is a complex number.
For $t=1$, we have $z^{2}-1=0$
and at $t=0$ we recover our original problem. The problem of getting
all solutions of the original problem now reduces to tracking solutions
of $H(z,t)=0$ from $t=1$ where we know the solutions, i.e., $z=\pm1$,
to $t=0$. The choice of $z^{2}-1$ in Eq.~(\ref{eq:one_var_homotopy}),
called the \textit{start system}, is made for the following reasons: this system
has the same number of solutions as the original problem and is easy
to solve. For multivariate systems, a clever choice of a start system
is essential in reducing the computation as we shall see.

We briefly mention the numerical methods used in path-tracking from $t=1$ to $t=0$.
One of the ways to track the paths is to solve a differential equation that is satisfied
along all solution paths, say $z_{i}^{*}(t)$ for the $i^{th}$ solution path,
\begin{equation}
\frac{dH(z_{i}^{*}(t),t)}{dt}=\frac{\partial H(z_{i}^{*}(t),t)}{\partial z}\frac{dz_{i}^{*}(t)}{dt}+\frac{\partial H(z_{i}^{*}(t),t)}{\partial t}=0.
\end{equation}
This equation is called the {\bf Davidenko differential equation}.
We can solve this initial value problem numerically -- again, pretending
that an exact solution is not known -- with the initial conditions as
$z_{1}^{*}(0)=1$ and $z_{2}^{*}(0)=-1$.  (Since we also know that $H(z_i^*(t),t) = 0$,
predictor-corrector methods, such as Euler's predictor and Newton's corrector,
are used in practice to solve this initial value problem.)
We shall not discuss the actual path tracking algorithms further,
but it is important to mention that in these algorithms are designed to handle
almost all apparent difficulties such as tracking singular solutions, multiple roots,
solutions at infinity, etc.

We now return to the complex number $\gamma$,
where we will consider $\gamma=e^{i\theta}$ with $\theta\in[0,2\pi)$ chosen generically.
It is known that for all but finitely values of $\theta\in[0,2\pi)$,
the paths are well-behaved for $t\in(0,1]$, i.e., for the whole path
except the end-point \cite{SW:95}.  This makes sure that there is no singularity
or bifurcation along the paths.  This is a remarkable technique, called
the {\bf $\gamma$-trick} in the literature, and constitutes the reason why
{\em NPHC is guaranteed to find all isolated solutions}.

There are several sophisticated numerical packages well-equipped with
path trackers such as {\sf Bertini} \cite{BHSW06}, {\sf PHCpack}~\cite{Ver:99},
{\sf PHoM}~\cite{GKKTFM:04}, {\sf HOMPACK}~\cite{MSW:89} and {\sf HOM4PS2}~\cite{GLW:05,Li:03}.
They all are available as freewares from the respective research groups.
In the above example, {\sf PHCpack} with its default settings yields
the solutions
$z= -2.23606797749979+i\,0.00000000000000$ and
$z = 2.23606797749979+i\,0.00000000000000$.
Thus, it gives the expected
two solutions of the system with a very high numerical accuracy.


\subsection{Multivariate Polynomial Homotopy Continuation}

We will now generalize the NPHC method explained above to the multivariate case,
say $P(x)=0$, where $P(x)=(p_{1}(x),\dots,p_{m}(x))$ and $x=(x_{1},\dots,x_{m})$, that
is \textit{known to have isolated solutions} (i.e., a $0$-dimensional ideal).
Of course, in many cases where one does not know the dimensionality of the solutions to start with, one has to check if the given system is an $0$-dimensional first. In the following sections, where we explain the extension of the NPHC method to the positive dimensional systems, we will describe a concrete way of finding if the system possesses only isolated solutions. One quick and dirty way of seeing that a square system (i.e., number of equations and number of variables are the same) has only isolated nonsingular solutions is to
add the determinant of the Jacobian matrix to the system and verifying that the combined system has no solutions. In the $0$-dimensional case, the system has
maximal rank at each solution.
Having said this, in many physical applications, we know {\em a priori} that the given system has only isolated solutions.

We can construct a homotopy, or a set of problems, similar to the aforementioned one-dimensional case, as
\begin{equation}
H(x,t)= (1-t) P(x) + \gamma t Q(x),
\end{equation}
where $Q(x)$ is a system of polynomial equations, $Q(x)=(q_{1}(x),\dots,q_{m}(x))$
with the following properties:
\begin{enumerate}
\item The solutions of $Q(x)=H(x,1)=0$ are known or can be easily obtained.
$Q(x)$ is called the \textit{start system} and the solutions are
called the \textit{start solutions}.
\item The number of solutions of $Q(x)=H(x,1)=0$ is equal to an estimated
number (or an upper bound) of the solutions for $P(x)=0$.
\item The solution set of $H(x,t)=0$ for $0<t\le1$ consists of a finite
number of smooth paths, called homotopy paths, each parameterized by
$t\in(0,1]$.
\item Every isolated solution of $H(x,0)=P(x)=0$ can be reached by some
path originating at a solution of $H(x,1)=Q(x)=0$.
\end{enumerate}
We can then track all the paths corresponding to each solution of
$Q(x)=0$ from $t=1$ to $t=0$ and reach $H(x,0) = P(x) = 0$. By implementing
an efficient path tracker algorithm, we can get all the isolated solutions
of a system of multivariate polynomials just as in the univariate
case. There are several upper bounds on the number of solutions of
the system $P(x)=0$, which yield alternatives to constructing $Q(x)$ and
the homotopy $H(x,t)$ for the multivariate case. The ones that we will consider are the
(1) Total Degree Homotopy, (2) 2-Homogeneous Homotopy, and (3) Polyhedral Homotopy.
We introduce the Total Degree Homotopy below and discuss the 2-Homogeneous and Polyhedral Homotopies in Appendices.
\subsubsection{Total Degree Homotopy}
The Total Degree Homotopy arises from the \textit{Classical B\'ezout Theorem}, which asserts that
a system of $m$ polynomial equations in $m$ variables has at most $\prod_{i=1}^{m}d_{i}$
isolated solution in $\mathbb{C}^m$, where $d_{i}$ is the degree of the $i$th
polynomial.  This bound, called the Classical B\'ezout Bound (CBB), is
sharp for generic values (i.e., roughly speaking, non-zero random
values) of the coefficients. The \textit{genericity} is well-defined and
the interested reader is referred to \cite{SW:95} for details.
The homotopy constructed using the CBB is called the \textit{Total
Degree Homotopy}.  The start system $Q(x)=0$ can be taken for example as
\begin{equation}
Q(x)=\left(\begin{array}{c}
x_{1}^{d_{1}}-1\\
x_{2}^{d_{2}}-1\\
\vdots \\
x_{m}^{d_{m}}-1
\end{array}\right)=0,\label{eq:Total_Degree_Homotopy}
\end{equation}
where $d_{i}$ is the degree of the $i^{th}$ polynomial of the original
system $P(x)=0$. Eq.~(\ref{eq:Total_Degree_Homotopy}) is easy to
solve and guarantees that the total number of start solutions is $\prod_{i=1}^{m}d_{i}$,
all of which are nonsingular. The Total Degree Homotopy is a very
effective and popular homotopy which is used in actual path trackers.

The advantages of the Total Degree Homotopy are (1) the CBB is easy
to compute, and (2) the start system based on the CBB can be solved quickly.
The drawback of it is that the CBB does not take the
sparsity of the system into account: systems arising in practice
have far fewer solutions than the CBB, so a large portion
of the computational effort is wasted.

The 2-Homogeneous Homotopy is constructed by first writing $\mathbb{C}^m = \mathbb{C}^k\times\mathbb{C}^{k-m}$
for some $0 < k < m$, which is accomplished by partitioning the original variables into
two groups.  This has the advantage of incorporating some of the structure of
the given polynomial system $P(x)$ into the start system $Q(x)$.
The corresponding bound, called the 2-Homogeneous B\'ezout Bound (2HomBB), is often
tighter than the CBB when the polynomial system $P(x)$ has a naturally arising
partition of the variables, which occurs in the examples below.
Given a partition, the 2HomBB is easy to compute and the start system can be
solved quickly via linear algebra.

The Polyhedral Homotopy uses the monomial structure of the given polynomial system $P(x)$
based on the Bernstein-Khovanskii-Kushnirenko (BKK) Theorem~\cite{Bernstein75,Khovanski78,Kushnirenko76}
to yield the BKK bound. Essentially, this upper bound on the number of complex
solutions is obtained by computing the mixed volume of the convex hull
of the Newton polytope (which is based on the exponents of the monomials appearing) of each equation.
Since we must introduce some jargon to fully describe the 2-Homogeneous and Polyhedral Homotopies,
we do it in Appendix \ref{app:2HomB} and \ref{app:polyhedral}, respectively. We note that, as with the CBB, the 2HomBB and BKK bound are also generically
sharp with respect to the family of polynomial systems under consideration.

Since the BKK count utilizes sparsity of the system, the BKK bound is tighter than the CBB and 2HomBB
resulting in fewer homotopy paths to track.
The apparent drawback is that to compute the BKK bound and solve the start system itself,
the Polyhedral Homotopy requires additional computational effort.
As we will see, the difference between the CBB and BKK bound suppress
this drawback of the Polyhedral Homotopy by saving a large portion
of the computational effort over the Total Degree Homotopy. The relationship
between the 2HomBB and BKK count vary based on the system.

\subsubsection{Parallelizability of the NPHC Method}

While going through the algorithm for the NPHC method, one could have noticed that to track paths from $t=1$ to $t=0$, i.e., from the start system
to the original system, each start solution does not need any knowledge about any other path.
In other words, each solution path can be tracked completely independently of all others.
This feature makes the method {\textit embarrassingly} parallelizable.
This is sharply different than the BA which is known to be a sequential algorithm, i.e.,
each step requires the knowledge of the previous step.
Thus, the NPHC method is destined to be much more efficient to solve bigger systems
which are way beyond the reach of the Gr\"obner basis techniques.

\subsubsection{Numerical Solutions of Multivariate System}
While tracking all the paths using either Total Degree, 2-Homogeneous, or Polyhedral Homotopies,
only those paths leading to solutions of the original system will converge at $t=0$, that is, some of the paths will diverge.
The values of the variables at $t=0$, i.e., $\vec{x}(t=0)$, after the path-tracking are the solutions of the original system.
Obviously, they will be in the numerical form. It is useful to define what we mean by a solution here.
For the multivariate case, a solution is a set of numerical values
of the variables which is within a given tolerance $\triangle_{\mbox{sol}}$ (we will take $\sim10^{-10}$ in our ensuing calculations)
of an actual solution.
Since the variables are allowed to take complex values, all the solutions
come with real and imaginary parts. A solution is a real solution
if the imaginary part of each of the variables is less than or equal
to a given tolerance, $\triangle_{\mathbb{R}}$ ($\sim10^{-7}$ is
a robust tolerance for the equations we will be dealing with in the
next section, below which the number of real solutions does not change).
All these solutions can be further refined to within an \textit{arbitrary
precision} (up to the memory and computational limits placed on by the machine).

The obvious question at this stage would be if the number of real
solutions depends on $\triangle_{\mathbb{R}}$. To resolve this issue,
we use a recently developed algorithm called ``alphaCertified'' which
is based on the so-called {\bf Smale's $\alpha$-theory}.  This algorithm
certifies which nonsingular solutions of the given polynomial system are
real using either exact rational arithmetic and arbitrary precision floating point
arithmetic \cite{2010arXiv1011.1091H}. This is a remarkable step,
because using alphaCertified, we can prove that a solution classified
as a real solution is actually a real solution independent of $\triangle_{\mathbb{R}}$,
and hence these numerical approximations are as good as the \textit{exact solutions}.

\section{Isolated Solutions and Phenomenology}\label{s:0dim}
We will now apply the principles introduced in the preceding section on
NPHC to questions which arise in theoretical physics, especially in string phenomenology
and supersymmetric gauge theory.
Our discussions will roughly fall into two categories to which we alluded in the discussions
above: (1) polynomials systems which have any isolated solutions; these are
clearly of importance when dealing with issues such as finding and stabilizing vacua
given some effective potential, and (2) systems which can have solution components;
these are crucial to understanding various geometrical properties of gauge theories such as
the moduli space of vacua or Calabi-Yau spaces in the AdS/CFT context.
We will first discuss (1) and then turn to (2) in the next section, illustrating in each case with
ample examples, particularly those which defy more standard techniques.

Much of current research in string phenomenology is focused on developing
methods to find and analyze vacua of four dimensional effective theories
for supergravity descended from flux compactifications.
Stated in explicit terms, one is interested in finding all the vacua (usually,
isolated stationary points) of the scalar potential $V$ of such a
theory. In particular, given a K\"ahler potential $K$, and a superpotential
$W$, for uncharged moduli fields, the scalar potential can be obtained as
\begin{equation}
V=e^{K}[\mathit{\mathit{\mathit{K^{A\bar{B}}}}}\ D_{A}W\ D_{\bar{B}}\bar{W}-3|W|^{2}] \ ,
\end{equation}
where $D_{A}$ is the K\"ahler derivative $\partial_{A}+\partial_{A}K$
and $\mathit{K}^{A\bar{B}}$ is the inverse of $\mathit{K_{A\bar{B}}}=\partial_{A}\partial_{\bar{B}}K$.
Once the vacua are found, one can then classify them by either using
the eigenvalues of the Hessian matrix of $V$ or by introducing further
constraints such as $W=0$.

Finding all the stationary points of a given potential $V$ amounts
to solving the stationary equations, i.e. solving the system of equations
consisting of the first derivatives of $V$ with respect to all the
fields equated to zero. The stationary equations for $V$ arising
in the string phenomenological models are usually nonlinear. In the
perturbative limit, $W$ usually has a polynomial form. This is an
important observation since we can then use algebraic geometric
concepts and methods to extract information about $V$.

For the systems known to have only isolated solutions, we are in the situation of $0$-dimensional
ideals, a Gr\"obner basis using a lexicographic ordering of the monomials
always has at least one univariate equation and the subsequent equations consist of an
increasing number of variables,
i.e., it is in a \textit{triangular form}.  The solutions of a Gr\"obner
basis is always the same as the original system, but the former is
easier to solve due to its triangular form as the univariate equation
can be solved either analytically or numerically using straightforward methods.
Then, by back-substituting the solutions in the subsequent equations
and continually solving them, we can find all the solutions of the original system.

\subsection{Some Toy Examples}
One of the main features of the ``{\sf Stringvacua}'' package \cite{Gray:2006gn,Gray:2008zs,Gray:2007yq} is to address
problems with a $0$-dimensional ideal using Gr\"obner basis techniques.
Let us first take some of the toy examples discussed therein and see how our numerical
techniques can extract some of the requisite information without resorting to the often
expensive Gr\"obner algorithms.

\paragraph{Sys1: A Single-Modulus Example:}
We begin with a single-modulus toy example.
Let the K\"ahler potential $K$ and superpotential $W$ be
given as
\begin{equation}
K=-3\log(T+\bar{T}) \ ,
\quad
W=a+bT^{8} \ .
\end{equation}
Note that the field $T$ comes
along with its complex conjugate. So even though they can be treated
as different variables by merely relabeling them, they are not actually
independent variables. To avoid this problem, we can write them in
terms of real and imaginary parts, i.e., $T=t+i\,\tau$ where $t$ and $\tau$ are real.
The potential is
\begin{equation}
V =  \frac{1}{3t}(4b(5b(t^{2}+\tau^{2})^{7}-3a(t^{6}-21t^{4}\tau^{2}+35t^{2}\tau^{4}-7\tau^{6})))
\end{equation}
which has $2$ variables. To find the stationary points of $V$,
we need to compute the zero locus of the partial
derivatives of $V$ with respect to variables $t$ and $\tau$:
\begin{eqnarray}
\nn
\frac{\partial V}{\partial t} & = & \frac{1}{3t^{2}}(4b(5b(13t^{2}-\tau^{2})(t^{2}+\tau^{2})^{6}-3a(5t^{6}-63t^{4}\tau^{2}+35t^{2}\tau^{4}+7\tau^{6})))=0,\\
\frac{\partial V}{\partial\tau} & = & \frac{1}{3t}(56b\tau(5b(t^{2}+\tau^{2})^{6}+a(9t^{4}-30t^{2}\tau^{2}+9\tau^{4})))=0 \ .
\end{eqnarray}
For general values of parameters $a$ and $b$ it was already known \cite{Gray:2006gn,Gray:2009fy} that the system becomes prohibitively
difficult to analyze using symbolic methods.  For now, we take $a=b=1$.

We also note that the stationary equations in this example involve
denominators. Since we are not interested in the solutions for which
the denominators are zero, we clear them out by multiplying them with
the numerators appropriately. In these equations, all the denominators
are multiples of $t$. The condition that none of the denominators
is zero can be imposed algebraically by adding a constraint equation
as $1-z\, t=0$ with $z$ being an additional variable. Thus there
are now $3$ equations in $3$ variables.
One could impose this by the method of saturation \cite{Gray:2006gn}.

{\sf Stringvacua} can easily solve this system: this system has $6$ real solutions.
Note that this computation uses the Gr\"obner basis technique implemented in {\sf Singular}.

We now turn to solve the above system using {\sf Bertini} and {\sf HOM4PS2}.
The CBB of this system is $364$, while the 2HomBB, using $\{t,\tau\}\times\{z\}$,
and the BKK bound are both $182$.  In the end, there are $56$ finite nonsingular complex solutions, of which
six are real.  Thus, we have produced the results known from symbolic methods
by using the NPHC method.  Note that the choice of $a$ and $b$ was crucial here.
For a generic choice for $a$ and $b$, the computation using symbolic methods via {\sf Stringvacua}
becomes extremely difficult already for this small system, whereas the NPHC method takes
the same time to solve it as before.  Since we have all the real solutions, we can then compute the Hessian
of $V$ at the real solutions and separate out the physically interesting vacua.
\paragraph{Sys2: Two-Moduli Model: }
Consider the K\"ahler potential and superpotential
\begin{eqnarray}
\nn K & = & -3\log(T+\bar{T})-\log(S+\bar{S}),\\
W & = & aS+bST+cT^{2},
\end{eqnarray}
with two fields $T=t+i\tau$ and $S=s+i\sigma$. Again, $a,b$ and $c$ are the parameters
and for convenience we chose them to be $1,-1,1$, respectively. The potential is
\begin{eqnarray}
\nn
V & = & \frac{1}{48st^{3}}(-5t^{4}+3(s^{2}+\sigma^{2})-2t^{2}\tau^{2}+3\tau^{4}-(s^{2}+\sigma^{2})(5t^{2}-3\tau^{2})-6(5st^{2}- \\
 &  & t(s^{2}+\sigma^{2})+2t\sigma\tau-s\tau^{2})+2(13st^{3}+t^{2}\sigma\tau+9st\tau^{2}-3\sigma\tau^{3}))
\end{eqnarray}
and the stationary equations are
\begin{eqnarray}
\nn
0 & = & \frac{1}{48st^{4}}(-5t^{4}-9(s^{2}+\sigma^{2})+2t^{2}\tau^{2}-9\tau^{4}+(s^{2}+\sigma^{2})(5t^{2}-9\tau^{2})\\
\nn
  &  & +6(5st^{2}-2t(s^{2}+\sigma^{2})+4t\sigma\tau-3s\tau^{2})-2\tau(t^{2}\sigma+18st\tau-9\sigma\tau^{2})),\\
\nn
0 & = & \frac{1}{48s^{2}t^{3}}(5t^{4}+3(s^{2}-\sigma^{2})+2t^{2}\tau^{2}-3\tau^{4}+6t(s^{2}-\sigma^{2}+2\sigma\tau)\\ \nn
 &  & -2\sigma\tau(t^{2}-3\tau^{2})-(s^{2}-\sigma^{2})(5t^{2}-3\tau^{2})),\\
\nn
0 & = & \frac{1}{24st^{3}}(-6t\sigma+t^{2}\sigma+6s\tau+18st\tau+3(s^{2}+\sigma^{2})\tau-9\sigma\tau^{2}\\
\nn
  &  & +2\tau(-t^{2}+3\tau^{2})),\\
0 & = & \frac{1}{24st^{3}}(3\sigma-5t^{2}\sigma+t^{2}\tau+3\sigma\tau^{2}-3\tau^{3}-6t(-\sigma+\tau)) \ .
\end{eqnarray}
 With the denominator equation $1-zst=0$, there are $5$ equations
in $5$ variables. The system possesses only isolated solutions and both
the {\sf Stringvacua} package and the NPHC method can solve this system. The CBB
is $432$, the 2HomBB, using $\{t,\tau,s,\sigma\}\times\{z\}$ is $144$,
and the BKK root count is $100$.  There are $6$ real solutions.

Again, for a generic choice of the values for $a,b$ and $c$,
it becomes extremely difficult for {\sf Stringvacua} to solve the corresponding system unlike the NPHC method.

\paragraph{Sys3-1: Two Moduli Example :}
We now take a four dimensional $\mathcal{N}=1$ supergravity theory with
K\"ahler potential and superpotential
\begin{eqnarray}
\nn
K & = & -3\log(T_{1}+\bar{T}_{1})-3\log(T_{2}+\bar{T}_{2}),\\
W & = & -T_{1}^{2}-T_{1}T_{2}-T_{2}^{2}+10(T_{1}+T_{2})-100.
\end{eqnarray}
Taking $T_{j}=t_{j}+i\tau_{j}$, for $j=1,2$, the potential is
\begin{eqnarray}
\nn
V & = & \frac{1}{192t_{1}^{3}t_{2}^{3}}(t_{1}^{4}+t_{1}^{3}(20-14t_{2})+20t_{2}^{3}+t_{2}^{4}-60t_{2}(100+2\tau_{1}\tau_{2}+\tau_{2}^{2})\\
\nn
  &  & +t_{1}^{2}(-500+280t_{2}-37t_{2}^{2}+10\tau_{1}^{2}+10\tau_{1}\tau_{2}+7\tau_{2}^{2})\\
\nn
  &  & +t_{2}^{2}(7\tau_{1}^{2}+10\tau_{1}\tau_{2}+10(-50+\tau_{2}^{2}))-2t_{1}(-140t_{2}^{2}+7t_{2}^{3}\\
 &  & +30(100+\tau_{1}^{2}+2\tau_{1}\tau_{2})+3t_{2}(200+\tau_{1}^{2}+4\tau_{1}\tau_{2}+\tau_{2}^{2}))\\
 \nn &  & +9(10000+\tau_{1}^{4}+2\tau_{1}^{3}\tau_{2}-100\tau_{2}^{2}+2\tau_{1}\tau_{2}^{3}+\tau_{2}^{4}+\tau_{1}^{2}(-100+3\tau_{2}^{2})))
\end{eqnarray}
with the stationary points dictated by
\begin{equation}\begin{array}{rcl}
\frac{\partial V}{\partial\tau_{1}} & = & \frac{1}{96t_{1}^{3}t_{2}^{3}}(-900\tau_{1}-60t_{1}\tau_{1}+10t_{1}^{2}\tau_{1}-6t_{1}t_{2}\tau_{1}+7t_{2}^{2}\tau_{1}+18\tau_{1}^{3}-60t_{1}\tau_{2}\\
 &  & +5t_{1}^{2}\tau_{2}-60t_{2}\tau_{2}-12t_{1}t_{2}\tau_{2}+5t_{2}^{2}\tau_{2}+27\tau_{1}^{2}\tau_{2}+27\tau_{1}\tau_{2}^{2}+9\tau_{2}^{3}),\\
\frac{\partial V}{\partial\tau_{2}} & = & \frac{1}{96t_{1}^{3}t_{2}^{3}}(-60t_{1}\tau_{1}+5t_{1}^{2}\tau_{1}-60t_{2}\tau_{1}-12t_{1}t_{2}\tau_{1}+5t_{2}^{2}\tau_{1}+9\tau_{1}^{3}\\
 &  &-900\tau_{2}+7t_{1}^{2}\tau_{2}-60t_{2}\tau_{2}-6t_{1}t_{2}\tau_{2}+10t_{2}^{2}\tau_{2}+27\tau_{1}^{2}\tau_{2}+27\tau_{1}\tau_{2}^{2}+18\tau_{2}^{3}),\\
\frac{\partial V}{\partial t_{1}} & = & \frac{1}{192t_{1}^{4}t_{2}^{3}}(-270000+12000t_{1}+500t_{1}^{2}+t_{1}^{4}+18000t_{2}+2400t_{1}t_{2}-280t_{1}^{2}t_{2}\\
&  & +1500t_{2}^{2}-560t_{1}t_{2}^{2}+37t_{1}^{2}t_{2}^{2}-60t_{2}^{3}+28t_{1}t_{2}^{3}-3t_{2}^{4}\\
 &  & +2700\tau_{1}^{2}+120t_{1}\tau_{1}^{2}-10t_{1}^{2}\tau_{1}^{2}+12t_{1}t_{2}\tau_{1}^{2}-21t_{2}^{2}\tau_{1}^{2}-27\tau_{1}^{4} \\
&  & +240t_{1}\tau_{1}\tau_{2}-10t_{1}^{2}\tau_{1}\tau_{2}+360t_{2}\tau_{1}\tau_{2}+48t_{1}t_{2}\tau_{1}\tau_{2}-30t_{2}^{2}\tau_{1}\tau_{2}\\
&  & -54\tau_{1}^{3}\tau_{2}+2700\tau_{2}^{2}-7t_{1}^{2}\tau_{2}^{2}+180t_{2}\tau_{2}^{2}+12t_{1}t_{2}\tau_{2}^{2}-30t_{2}^{2}\tau_{2}^{2}\\
&  & -81\tau_{1}^{2}\tau_{2}^{2}-54\tau_{1}\tau_{2}^{3}-27\tau_{2}^{4}),\\
\frac{\partial V}{\partial t_{2}} & = & \frac{1}{192t_{1}^{3}t_{2}^{4}}(-270000+18000t_{1}+1500t_{1}^{2}-60t_{1}^{3}-3t_{1}^{4}\\
&  &+12000t_{2}+2400t_{1}t_{2}-560t_{1}^{2}t_{2}+28t_{1}^{3}t_{2}+500t_{2}^{2}-280t_{1}t_{2}^{2}+37t_{1}^{2}t_{2}^{2}+t_{2}^{4}\\
&  &+2700\tau_{1}^{2}+180t_{1}\tau_{1}^{2}-30t_{1}^{2}\tau_{1}^{2}+12t_{1}t_{2}\tau_{1}^{2}-7t_{2}^{2}\tau_{1}^{2}-27\tau_{1}^{4}\\
 &  & +360t_{1}\tau_{1}\tau_{2}-30t_{1}^{2}\tau_{1}\tau_{2}+240t_{2}\tau_{1}\tau_{2}+48t_{1}t_{2}\tau_{1}\tau_{2}\\
 &  &-10t_{2}^{2}\tau_{1}\tau_{2}-54\tau_{1}^{3}\tau_{2}+2700\tau_{2}^{2}-21t_{1}^{2}\tau_{2}^{2}+120t_{2}\tau_{2}^{2}\\
 &  &+12t_{1}t_{2}\tau_{2}^{2}-10t_{2}^{2}\tau_{2}^{2}-81\tau_{1}^{2}\tau_{2}^{2}-54\tau_{1}\tau_{2}^{3}-27\tau_{2}^{4}).
\end{array}\end{equation}

We can easily clear out the denominators from each of the equations
and add a constraint equation $1-z\, t_{1}\, t_{2}=0$ representing
the fact that none of the denominators is zero. This combined system
contains only isolated solutions, but solving this system using the symbolic Gr\"obner basis
methods is a prohibitively difficult task \cite{Gray:2006gn}.

However, using the NPHC method, we can solve this system in less than a minute
on a standard desktop machine.  The CBB of this system is $432$,
the 2HomBB, using $\{t_1,\tau_1,t_2,\tau_2\}\times\{z\}$, and the BKK
root count are $144$. In the end, there are $70$ solutions, of which
$6$ are real.

\paragraph{Sys3-2:}
Furthermore, the equations for F-terms are
\begin{eqnarray*}
0 & = & 300-10t_{1}-t_{1}^{2}-30t_{2}+t_{1}t_{2}+3t_{2}^{2}-3\tau_{1}^{2}-3\tau_{1}\tau_{2}-3\tau_{2}^{2},\\
0 & = & 300-30t_{1}+3t_{1}^{2}-10t_{2}+t_{1}t_{2}-t_{2}^{2}-3\tau_{1}^{2}-3\tau_{1}\tau_{2}-3\tau_{2}^{2},\\
0 & = & -30\tau_{1}+2t_{1}\tau_{1}+3t_{2}\tau_{1}-30\tau_{2}+t_{1}\tau_{2}+6t_{2}\tau_{2},\\
0 & = & -30\tau_{1}+6t_{1}\tau_{1}+t_{2}\tau_{1}-30\tau_{2}+3t_{1}\tau_{2}+2t_{2}\tau_{2}
\end{eqnarray*}
The CBB and the BKK root count both are the same in this case, namely $16$.
Due to the lack of a good partition of the variables, the 2HomBB is actually
larger, namely $24$ for $\{t_1,t_2\}\times\{\tau_1,\tau_2\}$, than the CBB for example.
These equations have only $12$ solutions, of which $4$ are real.

\comment{
\todo{Not sure we need this example any more since it's not more complicated than
the previous one and nothing new is really demonstrated here}
\paragraph{Two-Moduli Example: }
\comment{from {\sf Stringvacua} (Don't know the name. It is the
'Numerical Approximation' in the Examples in Stringvacua)}
The K\"ahler potential and superpotential are

\begin{eqnarray*}
K & = & -\log(T+\bar{T})-3\log(S+\bar{S})\\
W & = & 0.5T+\pi S+T^{2}
\end{eqnarray*}
 We take $T=t+i\tau$ and $S=s+i\sigma$, so the potential is

\begin{eqnarray*}
V & = & -\frac{1.02808}{st}+\frac{-0.589049-1.9635t+\frac{0.392699\tau^{2}}{t}}{s^{2}}+\\
 &  & \frac{1}{s^{3}}(0.1875t^{2}+0.5625t^{3}-0.785398\sigma\tau-0.0625\tau^{2}+t(0.015625+0.625\tau^{2})\\
 &  & +\frac{0.61685\sigma^{2}+0.19635\sigma\tau+0.015625\tau^{2}+0.0625\tau^{4}}{t})
\end{eqnarray*}
 whose stationary equations are

\begin{eqnarray*}
\frac{\partial V}{\partial t} & = & \frac{1}{s^{3}t^{2}}(1.0280837917801413s^{2}+0.375t^{3}+1.6875t^{4}-0.6168502750680849\sigma^{2}\\
 &  & -0.19634954084936207\sigma\tau-0.015625\tau^{2}-0.0625\tau^{4}+\\
 &  & s(-1.9634954084936207t^{2}-0.39269908169872414\tau^{2})+t^{2}(0.015625+0.625\tau^{2})),\\
\frac{\partial V}{\partial s} & = & \frac{1}{s^{4}t}(1.0280837917801413s^{2}-0.5625t^{3}-1.6875t^{4}-1.8505508252042546\sigma^{2}\\
 &  & -0.5890486225480862\sigma\tau+t(2.356194490192345\sigma+0.1875\tau)\tau\\
 &  & -0.046875\tau^{2}-0.1875\tau^{4}+t^{2}(-0.046875-1.875\tau^{2})+s(1.1780972450961724t\\
 &  & +3.9269908169872414t^{2}-0.7853981633974483\tau^{2})),\\
\frac{\partial V}{\partial\tau} & = & \frac{1}{s^{3}t}(0.19634954084936207-0.7853981633974483t)\sigma\\
 &  & +\tau(0.03125+0.7853981633974483s-0.125t+1.25t^{2}+0.25\tau^{2})),\\
\frac{\partial V}{\partial\sigma} & = & \frac{1}{s^{3}t}(1.2337005501361697\sigma+0.19634954084936207\tau-0.7853981633974483\, t\,\tau)
\end{eqnarray*}

The {}``NumRoots'' command gives $7$ distinct roots%
\footnote{This is because the NumRoots command does not take into account the
constraint equation for the denominators. Can you check if the number
of real roots for the system with the denominator equation obtained
using a corresponding NumRoots command agrees with the NPHC method?%
}.

For the system without the denominator equation, there are components
of dimension $0$ and there is one component of degree $4$ of dimension
1.

For the system with the denominator equation, the CBB is $288$. The
number of solutions is $5$ out which there are 3 real solutions.
There are $283$ infinite solutions.
}

\subsection{Examples from Heterotic and M-Theory Effective Potentials}
Having warmed up with some toy examples and seen that the NPHC method can
extract information which is too difficult for Gr\"obner basis methods or for ordinary
numerical methods untuned for polynomial systems, let us see some ``real'' scenarios
from phenomenology.

\paragraph{Sys4: A Heterotic Example: }
Let us begin with a K\"ahler potential and superpotential taken from a heterotic compactification \cite{Gurrieri:2004dt,deCarlos:2005kh}:
\begin{eqnarray}
\nn K & = & -3\ln(T+\bar{T})-3\ln(Z+\bar{Z}),\\
W & = & i(\xi+ieT)+(\epsilon+ipT)Z+\frac{i}{2}(\mu+iqT)Z^{2}+\frac{1}{6}(\rho+irT)Z^{3} \ ,
\end{eqnarray}
where $\xi,r,\epsilon,q,\mu,p,\rho,e$ are parameters which satisfy
the constraint $\xi r-\epsilon q+\mu p-\rho e=0$.
We write $T=t+i\tau$ and $Z=z+i\zeta$, and take $\xi=-13,r=0,\epsilon=-4,q=2,\mu=2,p=1,\rho=5,e=-7$
to ensure that the solution space is zero-dimensional.

The critical points satisfy
\begin{equation}
{\small
\begin{array}{rcl}
0 & = & -\frac{1}{6912t^{4}z^{3}}(164268+10152z^{2}-4032tz^{3}+2268z^{4}\\
 &  & -252t^{2}z^{4}+360tz^{5}+675z^{6}+101088\xi-11232tz\xi+25380z^{2}\xi\\
 &  & +1620z^{4}\xi+40824\xi^{2}-1728tz\xi^{2}+9720z^{2}\xi^{2}-216t^{2}z^{2}\xi^{2}\\
 &  & +720tz^{3}\xi^{2}+2025z^{4}\xi^{2}+28836\xi^{3}+3240z^{2}\xi^{3}+7452\xi^{4}\\
 &  & +36t^{2}\xi^{4}+360tz\xi^{4}+2025z^{2}\xi^{4}+1620\xi^{5}+675\xi^{6}-8424z^{2}T\\
 &  & -216z^{4}T-4320z^{2}\xi T-1620z^{4}\xi T-25272\xi^{2}T-2160z^{2}\xi^{2}T\\
 &  & -7776\xi^{3}T-3240z^{2}\xi^{3}T-1944\xi^{4}T-1620\xi^{5}T+108z^{4}T^{2}\\
 &  & +1080z^{2}\xi^{2}T^{2}+972\xi^{4}T^{2}+1764(t^{2}+27T^{2})\\
 &  & -36(t^{2}(7z^{2}-\xi^{2})-3(z^{2}+9\xi^{2})T^{2})+84(-6t^{2}\xi\\
 &  & -6tz(-24+25z^{2}-12\xi-15\xi^{2})-27T(78+24\xi+6\xi^{2}+5\xi^{3}\\
 &  & +z^{2}(2+5\xi)+6\xi T)+6(t^{2}(11z^{2}+\xi^{2})+9(z^{2}+3\xi^{2})T^{2}))\\
 &  & -36(2tz(-78+6\xi^{2}+10\xi^{3}+z^{2}(14+10\xi))-3(5z^{4}\\
 &  & +z^{2}(8+10\xi+20\xi^{2})+3\xi(78+24\xi+6\xi^{2}+5\xi^{3}))T\\
 &  & +2(t^{2}(-3z^{2}\xi+\xi^{3})+3\xi(5z^{2}+9\xi^{2})T^{2}))),\\
0 & = & -\frac{1}{2304t^{3}z^{4}}(54756+1128z^{2}-252z^{4}+84t^{2}z^{4}\\
 &  & -120tz^{5}-225z^{6}+33696\xi-3744tz\xi+2820z^{2}\xi-180z^{4}\xi\\
 &  & +13608\xi^{2}-576tz\xi^{2}+1080z^{2}\xi^{2}-72t^{2}z^{2}\xi^{2}\\
 &  & -225z^{4}\xi^{2}+9612\xi^{3}+360z^{2}\xi^{3}+2484\xi^{4}+36t^{2}\xi^{4}\\
 &  & +120tz\xi^{4}+225z^{2}\xi^{4}+540\xi^{5}+225\xi^{6}-936z^{2}T+24z^{4}T\\
 &  & -480z^{2}\xi T+180z^{4}\xi T-8424\xi^{2}T-240z^{2}\xi^{2}T-2592\xi^{3}T\\
 &  & -360z^{2}\xi^{3}T-648\xi^{4}T-540\xi^{5}T-12z^{4}T^{2}\\
 &  & +120z^{2}\xi^{2}T^{2}+324\xi^{4}T^{2}+1764(t^{2}+9T^{2})\\
 &  & +12(t^{2}(-7z^{2}+3\xi^{2})+(z^{2}+27\xi^{2})T^{2})\\
 &  & +12(-2tz(-78+6\xi^{2}+10\xi^{3})-(5z^{4}\\
 &  & +z^{2}(-8-10\xi-20\xi^{2})-9\xi(78+24\xi+6\xi^{2}+5\xi^{3}))T\\
 &  & +2\xi(3t^{2}(z^{2}-\xi^{2})-(5z^{2}+27\xi^{2})T^{2}))\\
 &  & +84(-6t^{2}\xi+6tz(8+\xi(4+5\xi))+2(t^{2}(11z^{2}+3\xi^{2})\\
 &  & +3(z^{2}+9\xi^{2})T^{2})-3T(72\xi+z^{2}(2+5\xi)\\
 &  & +3(78+6\xi^{2}+5\xi^{3}+6\xi T)))),\\
0 & = & \frac{1}{384t^{3}z^{3}}(-3276-240z^{2}-4z^{4} -1008\xi-290z^{2}\xi-30z^{4}\xi-720\xi^{2}-40z^{2}\xi^{2}-354\xi^{3}\\
 &  & -60z^{2}\xi^{3}-36\xi^{4}-30\xi^{5}+1764T+168z^{2}T+4z^{4}T\\
 &  & +504\xi^{2}T+40z^{2}\xi^{2}T+36\xi^{4}T+4(z^{2}+9\xi^{2})T\\
 &  & +2(5z^{4}+z^{2}(8+\xi(10+20\xi-20T))+3\xi(24\xi+5\xi^{3}\\
 &  & +3\xi^{2}(2-4T)-6(-13+14T)))),\\
0 & = & \frac{1}{1152t^{3}z^{3}}(5616-936tz+1410z^{2}\\

 &  & +90z^{4}+4536\xi-288tz\xi+1080z^{2}\xi-72t^{2}z^{2}\xi\\
 &  & +120tz^{3}\xi+225z^{4}\xi+4806\xi^{2}+540z^{2}\xi^{2}\\
 &  & +1656\xi^{3}+24t^{2}\xi^{3}+120tz\xi^{3}+450z^{2}\xi^{3}\\
 &  & +450\xi^{4}+225\xi^{5}-240z^{2}T-90z^{4}T-2808\xi T-240z^{2}\xi T\\
 &  & -1296\xi^{2}T-540z^{2}\xi^{2}T-432\xi^{3}T-450\xi^{4}T\\
 &  & +120z^{2}\xi T^{2}+216\xi^{3}T^{2}+12\xi(t^{2}+9T^{2})\\
 &  & -42(-12tz-4t^{2}\xi-30tz\xi+72T+15z^{2}T+36\xi T+45\xi^{2}T\\
 &  & -36\xi T^{2}+2(t^{2}+9T^{2}))+2(-3(2tz(5z^{2}+6\xi+15\xi^{2})\\
 &  & +(-144\xi-z^{2}(10+40\xi)-3(78+18\xi^{2}+20\xi^{3}))T)\\
 &  & +6(3t^{2}(z^{2}-\xi^{2})-(5z^{2}+27\xi^{2})T^{2}))).
\end{array}
}
\end{equation}
We can clear out the denominators from all the equations and add the
additional constraint equation $1-y\, z\, t=0$ with additional variable
$y$, to get a combined system which only possess of isolated solutions.
The system is prohibitively difficult to solve using the Gr\"obner
basis techniques.  However, the NPHC method solves it in less than
a minute. The CBB is $2700$, the 2HomBB using $\{t,\tau,z,\zeta\}\times\{y\}$ is $900$,
and the BKK root count is $340$.
There are a total of $62$ solutions, of which only $6$ are real.

\paragraph{Sys5: A Model from M-Theory: }

Here, we take an example of M theory compactified on the coset $\frac{\mbox{SU}(3)\times\mbox{U}(1)}{\mbox{U}(1)\times\mbox{U}(1)}$ \cite{Micu:2006ey}.
The coset has $\mbox{SU}(3)$ structure. The corresponding K\"ahler
and superpotential are
\begin{eqnarray*}
K & = & -4\log(-i(U-\bar{U}))-\log(-i(T_{1}-\bar{T}_{1})(T_{2}-\bar{T}_{2})(T_{3}-\bar{T}_{3})),\\
W & = & \frac{1}{\sqrt{8}}(4U(T_{1}+T_{2}+T_{3})+2T_{2}T_{3}-T_{1}T_{3}-T_{1}T_{2}+200).
\end{eqnarray*}
Using $T_{j}=\tau_j-it_{j}$, for $j=1,2,3$, and $U=y-ix$,
the potential is
\begin{eqnarray*}
V & = & \frac{1}{256t_{1}t_{2}t_{3}x^{4}}(40000+t_{3}^{2}\tau_{1}^{2}-400\tau_{1}\tau_{2}-4t_{3}^{2}\tau_{1}\tau_{2}+4t_{3}^{2}\tau_{2}^{2}+\tau_{1}^{2}\tau_{2}^{2}-400\tau_{1}\tau_{3}\\
 &  & +800\tau_{2}\tau_{3}+2\tau_{1}^{2}\tau_{2}\tau_{3}-4\tau_{1}\tau_{2}^{2}\tau_{3}+\tau_{1}^{2}\tau_{3}^{2}-4\tau_{1}\tau_{2}\tau_{3}^{2}+4\tau_{2}^{2}\tau_{3}^{2}-24t_{2}t_{3}x^{2}\\
 &  & +4t_{3}^{2}x^{2}-24t_{1}(t_{2}+t_{3})x^{2}+4\tau_{1}^{2}x^{2}+8\tau_{1}\tau_{2}x^{2}+4\tau_{2}^{2}x^{2}+8\tau_{1}\tau_{3}x^{2}+8\tau_{2}\tau_{3}x^{2}\\
 &  & +4\tau_{3}^{2}x^{2}+1600\tau_{1}y-8t_{3}^{2}\tau_{1}y+1600\tau_{2}y+16t_{3}^{2}\tau_{2}y-8\tau_{1}^{2}\tau_{2}y-8\tau_{1}\tau_{2}^{2}y\\
 &  & +1600\tau_{3}y-8\tau_{1}^{2}\tau_{3}y+16\tau_{2}^{2}\tau_{3}y-8\tau_{1}\tau_{3}^{2}y+16\tau_{2}\tau_{3}^{2}y+16t_{3}^{2}y^{2}+16\tau_{1}^{2}y^{2}\\
 &  & +32\tau_{1}\tau_{2}y^{2}+16\tau_{2}^{2}y^{2}+32\tau_{1}\tau_{3}y^{2}+32\tau_{2}\tau_{3}y^{2}+16\tau_{3}^{2}y^{2}\\
 &  & +t_{1}^{2}(t_{2}^{2}+t_{3}^{2}+\tau_{2}^{2}+2\tau_{2}\tau_{3}+\tau_{3}^{2}+4x^{2}-8\tau_{2}y-8\tau_{3}y+16y^{2})\\
 &  & +t_{2}^{2}(4t_{3}^{2}+\tau_{1}^{2}-4\tau_{1}(\tau_{3}+2y)+4(\tau_{3}^{2}+x^{2}+4\tau_{3}y+4y^{2})).
\end{eqnarray*}
There are $9$ equations to be solved: $8$ equations corresponding to the derivatives with respect to $8$ variables plus one
constraint equation $1 - z(t_{1} t_{2} t_{3} x) = 0$ with $z$ being an additional variable.
The CBB is 103,680, the 2HomBB, using $\{t_j,\tau_j,x,y\}\times\{z\}$ is
20,736, and the BKK root count is 18,624.
The NPHC method yields $516$ solutions, of which only $12$ are real.
The computation based on the BKK root count takes around $10$ minutes
using {\sf HOM4PS2}.

\paragraph{Sys6: A $SU(2)\times SU(2)/\mathbb{Z}_{2}\times\mathbb{Z}_{2}$ Model: }

Compactifying type IIA supergravity on $SU(2)\times SU(2)/\mathbb{Z}_{2}\times\mathbb{Z}_{2}=S^{3}\times S^{3}/\mathbb{Z}_{2}\times\mathbb{Z}_{2}$
and restricting to the modes that are left-invariant under the action
of $SU(2)\times SU(2)$ one obtains an $\mathcal{N}=1$ supergravity
in 4D, which was studied in \cite{Danielsson:2011au,Caviezel:2008tf,Flauger:2008ad}.
There are no D-terms so that the scalar potential $V$ is determined
by the K\"ahler potential $K$ and the superpotential $W$ which are
\begin{align*}
K & =-\log((t_{1}+\bar{t}_{1})(t_{2}+\bar{t}_{2})(t_{3}+\bar{t}_{3}))\\
 & -\log(16(N_{1}+\bar{N}_{1})(N_{2}+\bar{N}_{2})(N_{3}+\bar{N}_{3})(N_{4}+\bar{N}_{4}))
\end{align*}
and
\begin{align*}
W & =-if_{0}t_{1}t_{2}t_{3}+t_{1}t_{2}f_{3}^{(2)}+t_{1}t_{3}f_{2}^{(2)}+t_{2}t_{3}f_{1}^{(2)}+(-i\, h+t_{1}+t_{2}+t_{3})N_{1}\\
 & +(-i\, h+t_{1}-t_{2}-t_{3})N_{2}+(i\, h+t_{1}-t_{2}+t_{3})N_{3}+(-i\, h-t_{1}-t_{2}+t_{3})N_{4}.
\end{align*}
Here $t_{i}$ are the K\"ahler moduli ($J-iB=t_{i}Y^{(2-)i}$), $N_{K}$
are the complex structure moduli ($e^{-\phi}Im(\Omega)+iC_{3}=N_{K}Y^{(3+)K}$),
$f_{0}$ is the mass parameter (of massive type IIA i.e. the flux
$F_{0}$), $f_{i}^{(2)}$ is the $F_{2}$-flux ($F_{2}=f_{i}^{(2)}Y^{(2-)i}$)
and $h$ is the $H$-flux (the $F_{4}$-flux is not turned on and
also included $O6$-planes).  The $Y$ are the elements of the cohomology of $M=SU(2)\times SU(2)/\mathbb{Z}_2\times \mathbb{Z}_2$.
The superscript $(i-)$ means that they are elements of $H^i(M,R)$ and the superscript $+/-$ means that they are even/odd under the orientifold projection. $i$ runs over the number of odd 2-forms in cohomology.
The fluxes and fields have fixed transformations under the orientifold projection and have to be expanded either even or odd cohomology in order to survive the orientifold projection.
By rescaling the fields and the overall
scale of $V$ we can set $f_{0},f_{1}^{(2)},f_{2}^{(2)},f_{3}^{(2)}$
to unity. Furthermore, the scalar potential simplifies substantially
if we analytically solve for the four $C_{3}$ axions (i.e. $Im(N_{K})$).
This leaves us with a model having
10 real fields and one parameter $h$.

As was shown in \cite{Caviezel:2008tf,Flauger:2008ad}, the scalar
potential $V$ for this model circumvents all known no-go theorems
against dS solutions and actually allows for dS critical points that
are however unstable. The results in \cite{Caviezel:2008tf,Flauger:2008ad}
where obtained by numerically minimizing $\epsilon=\frac{K^{A\bar{B}}(\partial_{A}V)\partial_{\bar{B}}V}{V^{2}}$
using Mathematica. This method however does not give all possible
solutions. Using {\sf HOM4PS2}, we have now solved $\partial_{A}V=0$ and
found all critical points for any given value of $h$. There are $10$
variables and $10$ equations, plus an additional constraint equation with additional variable
to make sure that the denominator is non-zero, yielding a total of
$11$ equations.  The CBB for this system is 279,936,000,
the 2HomBB, using the original variables in one group and the additional variable
in the other, is 34,992,000, but the BKK bound is only 574,080.
For a fixed value of $h(=-5)$ we have found that there are $60$ real solutions
to these equations. By stability analysis of the solutions, we conclude
that only one (the previously known) unstable dS critical point exists
in this case. Furthermore, we also find (previously unknown) that
up to permutation of the fields there are $4$ additional AdS solutions.
Thus, we have now completely classified the stationary points for
this model using the numerical homotopy continuation method.

\begin{table}
\begin{tabular}{|c|c|c|c|c|c|c|c|}
\hline
Model  & \# Eqns  & CBB & 2HomBB & BKK  & Real Sols  & Timing  & GB Technique\tabularnewline
\hline
Sys1  & $3$  & $364$  & $182$ & $182$  & $6$  & $<20$s  & Yes \tabularnewline
\hline
Sys2  & $5$  & $432$  & $144$ & $100$  & $6$  & $<10$s  & Yes\tabularnewline
\hline
\hline
Sys3-1  & $5$  & $432$  & $144$ & $144$  & $6$  & $<30$s  & No\tabularnewline
\hline
Sys3-2 & $4$  & $16$  & $24$ & $16$  & $4$  & $<10$s  & Yes\tabularnewline
\hline
Sys4  & $5$  & $2700$  & $900$ & $340$  & $6$  & $<10$s  & No\tabularnewline
\hline
Sys5  & $9$  & $103680$ & $20736$ & $18624$ & $12$ & $\sim590$s  & No\tabularnewline
\hline
Sys6  & $11$ & $279936000$ & $34992000$ & $574080$ & $60$ & $11$ hrs & No\tabularnewline
\hline
\end{tabular}\caption{The short names of the various models used in the main text are written in the first column. The subsequent columns are the number of equations,
CBB bound, 2HomBB bound, BKK bound, number of real solutions, and the time taken by the NPHC method (either {\sf Bertini} or {\sf HOM4PS2}). The final column lists whether or not the Gr\"obner Basis technique (via {\sf Stringvacua}) could solve the corresponding system.}
\end{table}

\section{Numerical Algebraic Geometry}
\label{s:nag}
We have presented, in the above, extensive examples wherein the power
of numerical algebraic geometry may be harnessed in various phenomenological
contexts, especially in the identification of isolated extrema of potentials
coming from heterotic and M-theory scenarios. One might naturally
wonder as to situations where the solution space of algebraic systems
has positive dimensionality. Such cases arise naturally as a central
subject in the study of supersymmetric gauge theories.

Indeed, the vacuum moduli space, parameterized by the scalar components
of multiplets, in generic gauge theories with supersymmetry is a continuous
manifold, or, more strictly speaking, an affine algebraic variety.
This fact is particularly pronounced in string theory when the gauge
theory manifests as the world-volume theory on a brane and the vacuum
moduli space is some space of special holonomy such as Calabi-Yau
or $G_{2}$ manifolds: this is the AdS/CFT correspondence.

The geometrical engineering of world-volume gauge theories given an
affine (Calabi-Yau) geometry has been a vast subject over the last
decade (cf.~e.g.~\cite{He:2004rn} for an introduction) and conversely,
the geometric analysis of the vacuum of a given supersymmetric gauge
theory (whether it comes from string theory or not) could uncover
hidden phenomenological symmetries \cite{Gray:2005sr,Gray:2006jb,Gray:2008yu}.
The algebraic geometry of the vacuum could often be extremely complicated
since the coordinates parameterizing it are the gauge invariant operators,
subject to superpotential constraints. Though in AdS/CFT, the geometries
are affine Calabi-Yau threefolds by construction, reverse engineering
\cite{Berenstein:2002ge} can lead to unlimited possibilities of K\"ahler
manifolds. In super-symmetric QCD with gauge group $SU(N_{c})$ with
$N_{f}$ fundamental flavours, for example, the vacuum is an affine
variety of dimension as high as $2N_{c}N_{f}-(N_{c}^{2}-1)$ when
$N_{f}\ge N_{c}$.

Furthermore, the vacuum moduli space could be composed of (possibly
intersecting) unions of components - or \textit{branches} - at various
dimensions all the way from 0 to that of the top (or coherent) component.
This phenomenon is especially marked in the investigation of the so-called
{}``master space'', which is a join of mesonic and baryonic branches
\cite{Forcella:2008bb,Forcella:2008eh}.

The importance of studying the algebraic geometric structure of the
vacuum, as can be seen from the above-mentioned, should not be undermined
by the technical difficulties which we are inevitably lead to encounter.
The usual method of attack is to recast the algebraic equations describing
the vacuum as a polynomial ideal, transform into standard Gr\"obner
basis, and then perform primary decomposition to extract the irreducible
branches. As is by now well recognized, this is a very expensive computation,
given especially the double-exponential running time of computing
Gr\"obner bases (cf.~\cite{Gray:2009fy}). In this section, we will
study examples where if one wishes to know simply some crucial but
preliminary information such as the number, dimension and degree of
the components, one could bypass the prohibitive step of primary and irreducible decomposition
and turn to the virtues of numerical algebraic geometry.

For positive dimensional varieties, since there are infinitely many
solutions, one needs a proper representation of the solutions.  The
important first question to be asked is how we need to represent the
solutions. For a $0$-dimensional variety, the solutions are just
a finite set of points, so they can be represented by complex numbers.
For positive dimensional varieties the situation is more involved.
The solutions in this case form, for example, curves, surfaces, or hypersurfaces.
A way to represent the solutions is to compute a parameterization of
the varieties, which can be accomplished using a Gr\"obner basis.
However, this is computationally very expensive.
The reader can find a nice discussion about this in \cite{CLO:07}.

Numerical Algebraic Geometry (NAG), on the other hand, cleverly uses
another approach in which the solutions are represented as Witness
Sets. We start with the fact that the number of points at which
an irreducible component, say of dimension $d$ in $\mathbb{C}^n$,
of a system of polynomial equations intersects a random
linear space of dimension $c=n-d$ is equal to the degree of the component. For example,
a cubic curve in $\mathbb{C}^3$ intersects with a two-dimensional
random hyperplane in three points. These intersection points
are called Witness points.  A \textit{random} $c$-dimensional linear space
means the solution set of $d=n-c$ linear equations with random coefficients.
Here, each linear equation is of the form $\sum_{j=1}^n C_j x_j = C_0$
where $C_k$ is a random complex number.
After computing the Witness points, one can slide this random
linear space around to obtain as many points on the component as
needed. This yields a \textit{parameterization} of the component,
called the Witness sets representation.

Now, it should be clear how the NPHC method plays an important role here.
Algebraically, we add $d$ linear equations to the system of equations and then
compute the isolated solutions of the combined system using the NPHC method.
The number of isolated solutions gives the number of Witness points for dimension $d$, i.e.,
the degree of the union of components of dimension $d$.
One can do this for all $d$ between $0$ and $n$ to compute the Witness points for all dimensions.
We note that, in practice, a sequence of homotopies, called a cascade \cite{RegenCascade,Cascade},
is used to compute the Witness points for all dimensions.

Given the Witness points for the components of dimension $d$, one needs to partition these points
into Witness points for the irreducible components.  This is done by the so-called monodromy algorithm.
The basic approach is to consider the Witness points as the linear space is moved around and then back
to the original linear space. In this move, some of the Witness points return to the same point.  However,
if a Witness point returns to a different point, these two points must be on the same component and
thus are grouped together. This process is repeated until each group is verified to be the set of Witness points
for an irreducible component. Since the full technical description of this verification step
of monodromy is beyond the scope of this article, we refer the reader to \cite{SW:95} for more details.

A simple example to consider is that of the affine twisted cubic curve which is given by the equations:
\begin{equation}
x^2 - y = 0, \, x^3 - z = 0.
\end{equation}
Running {\sf Bertini}, for example, on these equations yields that the irreducible decomposition of
the variety contains one component of dimension $1$ and degree $3$, as expected. It also gives a
random hyperplane that was used to slice the variety so that we can then get as many points as
we like on this positive dimensional component.

\subsection{The Master Space}
First we apply NAG to a model where the results are at least partially known, so that we can have a bench-marking to our NAG set up. In \cite{Forcella:2008bb,Forcella:2008eh}, we introduced the concept
of the {}``master space'' of supersymmetric gauge theories, which
controls the combined mesonic and baryonic branches of the vacuum
moduli space. Computationally, this is the space of F-flatness, that
is, the Jacobian ideal of the superpotential and can be construed
as a baryonic fibration over the mesonic moduli space. In string theory,
in the situation of a single D3-brane probing an affine toric Calabi-Yau
threefold, the resulting world-volume physics is a $U(1)^{g}$ quiver
gauge theory; here, the master space is a toric variety of dimension
$g+2$, whose K\"ahler quotient by the $g-1$ independent $U(1)$-actions
is the Calabi-Yau threefold.

Though in the toric case, polytope and plethystic techniques can simplify
the computation of the master space, direct attack thereon is still
needed in general. Once again, for large number of fields, recasting
the Jacobian ideal into standard bases may become prohibitively difficult
and numerical methods can become a more natural choice in quickly
extracting the dimension and irreducible components of the master
space.

We will exemplify with $\cN=1$ gauge theories for a D3-brane on the
Abelian orbifold $\IC^{3}/\IZ_{m}\times\IZ_{n}$. This is a toric
case which had been approached by both direct and combinatorial methods
for some low values of $(m,n)$ in \cite{Forcella:2008bb}, and hence
provide a good point of reference. Another advantage we have here
is that the Jacobian ideal is always {}``square'', in the sense
that the number of variables is equation to the number of vanishing
equations. To summarize, the gauge theory of interest is a
quiver theory with $mn$ nodes, a total of $3mn$ bifundamental fields
$\{X_{i,j},\ Y_{i,j},\ Z_{i,j}\}$ from node $i$ to node $j$ (with
$(i,j)$ defined modulo $(k,m)$ and with superpotential
\begin{equation}
W=\sum\limits _{i=0}^{k-1}\sum\limits _{j=0}^{m-1}X_{i,j}Y_{i+1,j}Z_{i+1,j+1}-Y_{ij}X_{i,j+1}Z_{i+1,j+1}\ .
\end{equation}

In Table \ref{table:m_n}, we provide the complete set of results for the
systems up to $(m,n)=(3,3)$. In \cite{Forcella:2008bb}, the top dimensional components of these
systems were computed using toric variety methods. Our results agree with the results
in this reference. In addition, we can now compute the complete irreducible decomposition of all these systems hence already for this simple set of systems we have got new and complete results.

For example, with $(m,n)=(2,2)$, we arrive at a master space which
is of dimension 16, in agreement with \cite{Forcella:2008bb}. We
further find that the variety reduces into 4 components: a
dimension 6 piece of degree 14 and 3 linear pieces (i.e., degree 1) of dimension 4.

One can also go further than these systems.  For example,
(4,1): 1 component of dimension 6 and degree 8; 33 components of dimension
4 and degree 1, (5,1): 1 component of dimension 7 and degree 16; 131
components of dimension 5 and degree 1, and (6,1): 1 component of dimension
8 and degree 32; 473 components of dimension 6 and degree 1.
For bigger systems such as (4,2), the decomposition space is more interesting, i.e., it has 1 component of dimension 10 and degree 584, 9 components of degree 1 and dimension 8, 8 components of degree 3 and dimension 8, 8 components of degree 4 and dimension 8, 24 components of degree 9 and dimension 8, and 8 components of degree 2 and dimension 6.

For (5,2), there is 1 component of degree 3632 and dimension 12, 17 component of degree 1 and dimension 10, 10 components of degree 3 and dimension 10, 40 components of degree 4 and dimension 10, 40 components of degree 9 and dimension 10, 40 components of degree 16 and dimension 10, 80 components of degree 27 and dimension 10, 130 components of degree 2 and dimension 8, 20 components of degree 6 and dimension 8, and 20 components of degree 8 and dimension 8. We can go on for bigger and bigger systems, however, we would rather stop here and move towards a more difficult system next.

\begin{table}
\begin{center}
\begin{tabular}{|c|c|c|c|c|}
\hline
$m$\textbackslash{}$n$ & 1       & 2              & 3               \tabularnewline
\hline
\hline
1  & NA           & $(1;4;2|6)$,   & $(1;5;4|9)$,     \tabularnewline
   &              & $(1;2;1|6)$    & $(7;3;1|9)$      \tabularnewline
\hline
2  & $(1;4;2|6)$, & $(1;6;14|12)$  & $(1;8;92|18)$,   \tabularnewline
   & $(1;2;1|6)$  & $(3;4;1|12)$   & $(5;6;1|18)$      \tabularnewline
   &              &                & $(6;6;3|18)$       \tabularnewline
\hline
3  & $(1;5;4|9)$, & $(1;8;92|18)$  & $(1;11;1620|27)$,  \tabularnewline
   & $(7;3;1|9)$  & $(5;6;1|18)$,  & $(6;9;1|27)$,   \tabularnewline
   &              & $(6;6;3|18)$   & $(27;9;2|27)$,   \tabularnewline
   &              &                & $(36;9;7|27)$,   \tabularnewline
   &              &                & $(27;7;1|27)$   \tabularnewline
\hline
\end{tabular}
\end{center}
\caption{The master space $F^{b}$ for $\mathbb{C}^{3}/\mathbb{Z}_{m}\times\mathbb{Z}_{n}$
as explicit varieties for several values of $m$ and $n$. $(p;q;r|d)$
means $p$ components of dimension $q$ and degree $r$ contained in
$d$-dimensional complex space.}\label{table:m_n}
\end{table}

\subsection{Supersymmetric Quantum Chromodynamics}

We now move to a more difficult problem of the familiar example of pure sQCD with gauge group
$SU(N_{c})$, $N_{f}$ number of fundamental flavours of quarks $Q$
and anti-quarks $\widetilde{Q}$, and no superpotential, so that the
matter content is summarized as:
\begin{equation}
{\scriptsize
\begin{tabular}{|c||c|cccccc|}
\hline
& \textsc{gauge symmetry} & & & \textsc{global symmetry} & & & \\
& $SU(N_c)$ & $SU(N_f)_L$ & $SU(N_f)_R$ & $U(1)_B$ & $U(1)_R$ & $U(1)_Q$ & $U(1)
_{\widetilde{Q} }$\\
\hline \hline
$Q^i_a$ & $\overline{\tiny\yng(1)}$ & $\tiny\yng(1)$ & $\mathbf{1}$ & 1 & $\frac
{N_f-N_c}{N_f}$ & 1 & 0 \\
$\widetilde{Q}^a_i$ & $\tiny\yng(1)$ & $\mathbf{1}$ & $\overline{\tiny\yng(1)}$
& -1 & $\frac{N_f-N_c}{N_f}$ & 0 & 1 \\
\hline
\end{tabular}
}
\end{equation}
The generators of the gauge invariant operators consists of the mesons and baryons:
\begin{equation}
\begin{array}{ll}
M^i_j = Q^i_a  \widetilde{Q}^a_j & \qquad \mbox{(mesons)} ~; \\
B^{i_1 \ldots i_{N_c}} = Q^{i_1}_{a_1} \ldots Q^{i_{N_c}}_{a_{N_c}} \epsilon^{a_1\ldots a_{N_c}} & \qquad \mbox{(baryons)} ~; \\
 \widetilde{B}_{i_1 \ldots i_{N_c}} = \widetilde{Q}^{a_1}_{i_1} \ldots \widetilde{Q}^ {a_{N_c}}_{i_{N_c}} \epsilon_{a_1 \ldots a_{N_c}} & \qquad \mbox{(antibaryons)} ~. \\
\end{array}
\end{equation}

It is a standard fact
that the dimension of the (classical) vacuum moduli space $\cM_{(N_{f},N_{c})}$
of the theory is
\begin{equation}
\dim\cM_{(N_{f},N_{c})}=\left\{ \begin{array}{lcl}
N_{f}^{2}\ , &  & N_{f}<N_{c}\ ;\\
2N_{c}N_{f}-(N_{c}^{2}-1)\ , &  & N_{f}\ge N_{c}\ .
\end{array}\right.
\end{equation}
More specifically, one can explicitly obtain the algebraic variety
\cite{Gray:2008yu} and prove they are, in fact, all affine Calabi-Yau
varieties (the reader is referred to Table 2 in \cite{Gray:2008yu}); for
example, $\cM_{(N_{f},N_{c})}$ is simply the
affine space $\IC^{N_{f}^{2}}$ when $N_{f}<N_{c}$, and, for $N_{f}=N_{c}$, it is a complete
intersection, with the Hilbert series given as
\begin{equation}
g^{N_{f}=N_{c}}(t)=\frac{1-t^{2N_{c}}}{(1-t^{2})^{N_{c}^{2}}(1-t^{N_{c}})^{2}}\ .
\end{equation}

In general, however, the vacuum is quite involved. Primary decomposition
to find the irreducible branches of the moduli space, as was experimented
in \cite{Gray:2008yu}, is prohibitively difficult, even for small
values of $N_{f}$ and $N_{c}$. It was conjectured, by working over
the coefficient fields of the rationals or number fields of finite
characteristic, that $\cM_{(N_{f},N_{c})}$ is actually irreducible.
Now, we can properly check the components by working over the complex
numbers, as one should. For example, take $(N_{f},N_{c})=(3,3)$,
we readily find that there is only one component, of complex dimension
10 and degree 3. For $(N_{f},N_{c})=(4,3)$, there is again only one
component of dimension $16$ and degree $115$.
A full study of the vacuum structure requires algebraic elimination, a subject which we will
address systematically in a forthcoming publication \cite{future}.

For now, we can extract information about the full mesonic moduli space by allowing Gr\"obner techniques to perform the elimination, such as using Macaulay2, and then use our numerical methods to perform the hard step of primary decomposition.
We tabulate some results in Table \ref{Table:N_f_N_c_eliminated}.
In \cite{Gray:2008yu}, only the top components of these systems using Macaulay2 were obtained and conjectured that the results were irreducible. In our case, we can now get the complete irreducible decomposition, not only the top dimension, of these systems. Hence our results yield that the conjecture was correct. We could have easily been able to go beyond the (4,4) systems, but the limiting factor was Macaulay2 itself which failed to obtain the eliminated ideals for the bigger systems for us.

\begin{table}
\begin{tabular}{|c|c|c|c|c|}
\hline
$N_{f}$\textbackslash{}$N_{c}$  & 1  & 2  & 3  & 4  \tabularnewline
\hline
\hline
1  & $(1;2;2|3)$  & - & - & - \tabularnewline
\hline
2  & $(1;4;6|8)$ & $(1;5;2|6)$  & - & - \tabularnewline
\hline
3  & $(1;6;20|15)$  & $(1;9;14|15)$  & $(1;10;3|11)$  & -  \tabularnewline
\hline
4  & $(1;8;70|24)$  & $(1;13;132|28)$ & $(1;16;115|24)$ & $(1;17;4|18)$ \tabularnewline
\hline
\end{tabular}
\caption{The irreducible decomposition of the moduli space of sQCD theories with $N_{f}$ fermions and $N_{c}$ bosons (after elimination from {\sf Macaulay 2}). Here, $N_f \geq N_c$, and the irreducible decomposition is presented in the form: $(\mbox{{no.components; dimension; degree | total dim}})$.
\label{Table:N_f_N_c_eliminated}}
\end{table}

Indeed, one could combine our present course of study with that of the previous
subsection. Indeed, the standard Yukawa term to sQCD for $N_{f}>N_{c}$,
in the above notation, is
\begin{equation}
W=\sum_{i,j=1}^{N_{f}}\sum_{a,b=1}^{N_{f}-N_{c}}\epsilon_{ab}M_{ij}Q_{a}^{i}\tilde{Q}_{b}^{j}\ .
\end{equation}
Here, $M_{ij}$ is the meson condensate of the quark-anti-quark.
This superpotential can be considered as being generated by Seiberg
duality from one with pure matter content.
For example, at $(N_{f},N_{C})=(3,2)$,
we find the master space to be of dimension 9, with 3 components,
one at degree 1 and two at degree 7. The complete results are shown in Table \ref{Table:N_f_N_c_uneliminated} and we see non-trivial primary components to the master space.
\begin{table}
\begin{tabular}{|c|c|c|c|c|}
\hline
$N_{f}$\textbackslash{}$N_{c}$  & 1  & 2  & 3  & 4  \tabularnewline
\hline
\hline
1  & $(1;0;1|2)$  & - & - & - \tabularnewline
\hline
2  & $(1;0;1|4)$  & $(1;5;4|8)$  & - & - \tabularnewline
\hline
3  & $(1;0;1|6)$  & $(1;7;6|12)$  & $(2;11;18|18)$  & -  \tabularnewline
\hline
4  & $(1;0;1|8)$  & $(1;9;8|16)$  & $(2;14;32|24)$ & (2;19;88|32), (1;20;320|32) \tabularnewline
\hline
\end{tabular}
\caption{The master space of sQCD theories with $N_{f}$ fermions and $N_{c}$ bosons. Here, $N_f \geq N_c$, and the irreducible decomposition is presented in the form: $(\mbox{{no.components; dimension; degree | total dim}})$.\label{Table:N_f_N_c_uneliminated}}
\end{table}

\subsection{Instanton Moduli Spaces}

The study of Yang-Mill instantons is, undoubtedly, another important
subject in the investigation of gauge theories. Ever since their discovery
by Belavin, Polyakov, Schwartz and Tyuplin \cite{Belavin:1975fg}
as well as the construction by Atiyah, Drinfeld, Hitchin and Manin
(ADHM) on self-dual solutions \cite{Atiyah:1978ri} in the 1970's,
the parameter, or moduli, space of these instanton solution has been
of great attention to physicists and mathematicians alike. The geometry
of the moduli space can be quite involved. Indeed, whereas the ADHM
construction gives the moduli space for the classical Lie groups,
that for the exceptional ones still remain a mystery.

With the embedding of the construction into string theory \cite{Witten:1994tz}
by Witten, Douglas and Moore, one could realize many cases as the
vacuum moduli space of the supersymmetric (quiver) theories which
we have introduced above. Recently, the algebraic geometry of the
one-instanton moduli space was analyzed along the line of our present
thought in \cite{Benvenuti:2010pq}, calculating, specifically, the
Hilbert series of the space. To clarify notation, we shall let $k$
$G$-instantons signify instantons of (classical) gauge group $G$
and with winding number $k$. Let us take Figure 7 of the said paper
as the illustrative starting point. Here, the moduli space of $k$
$SU(N)$ instantons is given as the $\cN=1$ vacuum moduli space of
the {}``flower quiver'', with the field content: $\phi^{(i=1,2)}$
and $\Phi$ charged as $k\times k$ matrix fields under $U(k)$, $X_{21}$
charged as $N\times k$ and $X_{12}$ as $k\times N$ matrix fields
respectively, all obeying the superpotential:
\begin{equation}
W=X_{21}\cdot\Phi\cdot X_{12}+\epsilon_{\alpha\beta}\phi^{(\alpha)}\cdot\Phi\cdot\phi^{(\alpha)}\ .
\end{equation}

Since the case of $k=1$ was considered in great detail in \cite{Benvenuti:2010pq},
let us move on the much more complicated and unsolved problem of,
say, $k=2$. We will focus on the Higgs branch by setting $\Phi$ to
have zero vacuum expectation value. The input data to the moduli space
is as follows.
\begin{equation}
\begin{array}{ll}
\mbox{Gauge invariants:}&\\
&\left(\tr(\phi^{(1)})^{a}(\phi^{(2)})^{b}\right)_{0<a+b\le2}\ ;\\
&\left(\sum\limits _{i=1}^{2}X_{12}^{i,j_{1}}X_{21}^{j_{2},i}\right)_{j_{1},j_{2}=1,\ldots,N}\ ;\\
&\left(\sum\limits _{i_{1},i_{2}=1}^{2}X_{12}^{i_{1},j_{1}}\phi_{i_{1},i_{2}}^{(\alpha)}X_{21}^{j_{2},i_{2}}\right)_{j_{1},j_{2}=1,\ldots,N,\alpha=1,2}\\
\mbox{F-Terms:}&\\
&0=\partial_{\mbox{fields}}\left|_{\Phi=0}\left[\tr X_{21}\cdot\Phi\cdot X_{12}-\tr\Phi[\phi^{(1)},\phi^{(2)}]\right]\right.\ .
\end{array}
\end{equation}
Let us be more specific and take, for example, at $N=2$. First,
the master space is easy to determine. It is a dimension 12 variety
of degree 16, defined by 4 quadrics in $\IC^{16}$. The Hilbert series
is
\begin{equation}
g(t;\cF_{(k,N)=(2,2)}^{\flat})=\frac{(1+t)^{4}}{(1-t)^{12}}\,\qquad PE^{-1}[g](t)=16t-4t^{2}\ .
\end{equation}
From the termination of the plethystic logarithm \cite{Benvenuti:2006qr},
we see that it is, as also suggested by the dimension, a complete
intersection.

This is an ideal with 21 generators in 33 variables:
16 variables in the $X$ and $\phi$ fields together with 17 auxiliary
variables.  It can be easily shown that this ideal has a unique irreducible
component of dimension 12 in $\mathbb{C}^{33}$, but computing its degree
and the actual moduli space via elimination are already quite overwhelming
for standard computer algebra packages.  {\sf Bertini},
in just under 24 hours using 200 processors in parallel, found that the
degree of this irreducible component of dimension 12 is 20364.
We will report on computing the moduli space via numerical elimination
in a forthcoming paper.

\section{A Few Words About NAG Packages}
Although it is not at all our intention to compare the performances of different packages in this paper, we convey
what the important computations are that different packages can do more efficiently than others.
There are so far three independently packages for NAG: {\sf Bertini}, {\sf HOM4PS2} and {\sf PHCPack}.

{\sf Bertini} is a general purpose package for NAG, i.e., it can find all the stationary points of a $0$-dimensional ideal
and also find irreducible decomposition of positive dimensional ideals efficiently. It also has many other
Numerical Algebraic Geometry implementations such as a membership test, finding multiplicities of isolated singular
solutions, facility to specify user-defined homotopy, etc.
In particular, {\sf Bertini} automatically constructs Total Degree and 2-Homogeneous Homotopies.
It utilizes adaptive multiprecision, and a parallel version of {\sf Bertini} is also publicly available.

{\sf HOM4PS2} is regarded as having the fastest path-tracker \cite{Li:03}.
It also has both Total Degree and Polyhedral Homotopies.
{\sf HOM4PS2} uses MixedVol-2.0 which is a very efficient package
to compute the BKK bound.
On the other hand, the applications of the current version of {\sf HOM4PS2} are limited: for the 0-dimensional case,
it only takes one of the two homotopies as its only input parameter along with the equations themselves and
gives out all the isolated real and complex solutions. For the positive dimensional systems, it also takes in only two of the
homotopies as its input parameter and the dimension of which one wants to get the irreducible decomposition of along with the
equations, and returns the irreducible decomposition of that particular dimension.
The parallel version of {\sf HOM4PS2} is under development and will be soon publicly available \cite{Li2009226}.

{\sf PHCPack} is a general purpose solver like {\sf Bertini}. It does quite a few things, such as
computing many different upper bounds on the number of solutions in addition to the CBB, 2HomBB, and BKK root counts,
it also can preprocess the system before running the path-tracker on it, one can supply a user-defined homotopy as well,
one can choose different numerical precision and parameters manually if one wants to, etc.
It runs interactively and one can chose different options in its menu-based interface.
The parallel version of {\sf PHCpack} is also publicly available.

All these packages have their own advantages. It is a goal of the authors of the current paper to write a separate
article on demonstrating various features of these packages specifically to the string phenomenologists and
particle physicists, and write a Mathematica interface for these packages in the same fashion as {\sf Stringvacua}.

\section{Conclusion and Prospects}\label{s:conc}
With the advances in computer algebra and algorithmic geometry, we have witnessed an
increasing trend over the last decade wherein such a paradigm has been gaining importance in
theoretical physics, especially in gauge and string theories.
Not only can hitherto unthinkable problems such as stabilizing highly complicated potentials or
scanning through huge classes of string vacua become feasible, but also new geometrical quantities which
characterize gauge theories such as Hilbert series or topological invariants of vacuum moduli spaces
can be calculated to elucidate the physics.

One great hurdle to this approach of computational algebraic geometry is that the central method
involved is finding the Gr\"obner basis of polynomial ideals, a highly non-parallelizable and exponential growth algorithm.
In this paper, we have demonstrated many instances where numerical algebraic geometry, especially
using homotopy continuation methods, can supplement this short-coming when addressing certain quantities.
Though by no means a replacement for the powers of the Gr\"obner techniques due to its symbolic nature, the numerical methods presented in this paper are
very useful and often computationally less expensive. In addition, the numerical methods are
parallelizable making it a spectacularly robust tool to address the string phenomenology related problems.

Two classes of problems, arising from a myriad of physical situations such as effective potentials in
string and M-theory compactifications and supersymmetric vacua of gauge theories, are particularly
amenable to our method: finding critical points of zero-dimensional ideals and irreducible decomposition of higher-dimensional polynomial ideals.
The former is a typical problem in vacuum stabilization problems and the latter, in extracting branches
of the moduli space of vacua.
We have shown many concrete examples where these are beyond current computational powers using Gr\"obner
bases but are fairly quickly done, in a highly parallelizable fashion, using numerical homotopy.

We are clearly only touching the surface of a reservoir of great utility.
The two classes of problems we have focused on in this paper, already of diverse applications, are
only two of the many quantities and techniques which have recently emerged to be crucial in the study of
gauge and string theories.
For example, the computation of the Hilbert series of a polynomial ideal, a problem seemingly of interest only
to pure algebraic geometry, has turned out to enumerate supersymmetric BPS spectra of operators
\cite{Benvenuti:2006qr}. Schubert calculus is another fascinating area coming from enumerative
geometry where the NAG method can be directly and efficiently applied.
A new homotopy meant to deal with the Schubert calculus problems has already been worked out.
Currently, methods for numerically obtaining the Hilbert series are being developed.
As another example, systematic elimination of variables in the so-called syzygy problem has turned to
be important in the study of moduli space of vacua in gauge theories and give explicit defining equations
of the vacuum manifold \cite{Gray:2005sr}. This, too, is being developed.
Indeed, there are several international collaborations which are engaged in the combing over the plethora of string vacua using computer search. Most of these ultimately reduce to manipulation of polynomial ideals and are thus limited in parallelizability and running time by the Gr\"obner paradigm.
The possibilities of numerically computing such sophisticated quantities as Hilbert series and ranks of (co-)kernels of polynomial maps should offer a new outlook.

\section*{Acknowledgments}
We are indebted to J.~Gray for many helpful discussions and collaboration in the initial phase of the project.
DM was supported by the US Department of Energy grant under contract no.
DE-FG02-85ER40237 and Science Foundation Ireland grant 08/RFP/PHY1462.
A large part of the paper is based on DM's ph.d. dissertation which was written in University of Adelaide and Imperial College London.
DM would like to thank all the group members of the {\sf Bertini}, {\sf HOM4PS2}, and {\sf PHCpack} packages for their constant help and support in technicalities all these years; and to Timm Wrase for pointing out and explaining the technical details of the $SU(2)\times SU(2)/\mathbb{Z}_{2}\times\mathbb{Z}_{2}$ model. YHH would like to thank the Science and
Technology Facilities Council, UK, for an Advanced Fellowship and grant
ST/J00037X/1, the Chinese Ministry of Education, for a Chang-Jiang
Chair Professorship at NanKai University, the US National Science Foundation for grant CCF-1048082,
as well as City University, London and Merton College, Oxford,
for their enduring support.  JDH would like to thank the US National Science Foundation
for grants DMS-0915211 and DMS-1114336.

\appendix

\section{2-Homogeneous Homotopy}
\label{app:2HomB}

As seen above, the variables in a polynomial system can often be partitioned
into two groups.  The key is a refined notion of the degree of a polynomial.
To illustrate, consider the polynomial $f(y,z) = 1 + 2y - 5z + 3y^2z$.
Clearly, the degree of $f$ is three.  However, $y$ has a maximum degree of $2$
and $z$ only appear linearly in $f$.  That is, if $\alpha$ and $\beta$ are
random constants and
\begin{equation}\label{eq:2HomConstant}
g(y) = f(y,\beta) \hbox{~~and~~} h(z) = f(\alpha,z),
\end{equation}
the degree of $g$ is $2$ while the degree of $h$ is $1$.
We say the {\bf bidegree} of $f$ is $(2,1)$.

The notion of the bidegree of $f(y,z)$ naturally extends to the case
when $y$ and $z$ are sets of variables.  That is, the bidegree of $f(y,z)$
is $(a,b)$ if the $\deg g = a$ and $\deg h = b$ where $g$
and $h$ are defined by (\ref{eq:2HomConstant}) with $\alpha$ and $\beta$
random vectors of the appropriate size.

For a system $P(x) = P(y,z)$ of $m$ polynomials in $m$ complex variables,
where $y$ is $k$ complex variables and $z$ is $m-k$ complex variables,
let $(a_i,b_i)$ be the bidegree of $P_i(y,z)$.
When each $P_i$ is nonconstant, the polynomial
\begin{equation}
R(s,t) = \prod_{i=1}^m (a_i s + b_i t)
\end{equation}
is homogeneous of degree $m$.
The 2-Homogeneous B\'ezout Theorem (see \cite{SW:95} for more details)
states that the number of isolated solutions
of $P(x) = 0$ in $\mathbb{C}^m$ is bounded above by the coefficient of the
$s^k t^{m-k}$ term of $R$.  This coefficient is the 2-Homogeneous B\'ezout bound (2HomBB).

For example, consider the system
\begin{eqnarray}
f_{1}(y,z) & = & 3y - 2z + 4z^3 = 0,\nonumber \\
f_{2}(y,z) & = & 1 - yz =0,\label{example_system_for_2hom}
\end{eqnarray}
where $y$ and $z$ are complex variables.
The CBB yields that this system has at most $3\cdot 2 = 6$ isolated solutions in $\mathbb{C}^2$.
The bidegree of $f_1(y,z)$ and $f_2(y,z)$ is $(1,3)$ and $(1,1)$, respectively, with
the polynomial $R(s,t) = (s + 3t)(s+t) = s^2 + 4st + 3t^2$.
Since the coefficient of $st$ in $R$ is $4$, the 2HomBB yields
that (\ref{example_system_for_2hom}) has at most $4$ isolated solutions in $\mathbb{C}^2$.

To construct a 2-Homogeneous Homotopy, we need to describe how to construct the
start system $Q(x) = Q(y,z)$.  The polynomials $P_i(y,z)$ and $Q_i(y,z)$ must have the same bidegree,
namely $(a_i,b_i)$.  One way to construct such a polynomial is to take
\begin{equation}\label{eq:2homQ}
Q_i(y,z) = \left(\prod_{j=1}^{a_i} L_{i,j}(y)\right)\cdot\left(\prod_{j=1}^{b_i} M_{i,j}(z)\right)
\end{equation}
where $L_{i,j}(y)$ and $M_{i,j}(z)$ are random linear polynomials.
The number of solutions of $Q(x) = Q(y,z) = 0$ is the 2HomBB of $P(x) = P(y,z)$,
and the solutions themselves can be computed using linear algebra.

\section{Polyhedral Homotopy}
\label{app:polyhedral}
As mentioned above, polynomial equations arising
in real-life problems are sparse in terms of the number of monomials which appear.
In general, CBB becomes an upper bound on the number of solutions for
such cases. We are looking for a tighter bound that takes
the sparsity of the system into account. There has been a huge amount
of work done on related issues using resultants and algebro-geometric methods,
but the most important result for us is Bernstein's theorem. In order
to state it clearly, we remind the reader of some standard notions of Laurent Polynomials,
Newton Polytopes, and Mixed Volume.

A Laurent polynomial allows negative exponents for the monomials,
so no variable is allowed to be zero. Hence, multiplication of any Laurent polynomial
by a monomial does not change the root count in $(\mathbb{C}^{*})^{m}=(\mathbb{C}/\{0\})^{m}$.
Formally, let $S_{i}\subset\mathbb{Z}^{m}$ be a set of vectors whose
elements are the exponents of the monomials of the $i$th polynomial.
$S_{i}$ is called the support of the $i$th polynomial. Then a polynomial
(say, $i$th polynomial) of the form $f_{i}(x)=\sum_{\alpha\in S_{i}}c_{i,\alpha}x^{\alpha}$
is called a Laurent polynomial. Here, $c_{i,\alpha}\in\mathbb{C}$
are the coefficients of the monomial $x^{\alpha}$ with $x\in(\mathbb{C}^{*})^{m}$.

Next, a set of points is called a convex set if for every pair of points
within the set (or more formally, the mathematical object made by
the set), every point on the straight line segment that joins them
is also within the set. The convex hull of a set X is the minimal
convex set containing X. We note that the convex hull of support $S_{i}$
of a polynomial, say $Q_{i}=\mbox{conv}(S_{i})$, is called the {\bf Newton
polytope} of $f_{i}(x)$.

For example, consider a two-variable system
\begin{eqnarray}
f_{1}(x,y) & = & 1+ax+bx^{2}y^{2}=0,\nonumber \\
f_{2}(x,y) & = & 1+cx+dy+exy^{2}=0,\label{example_system_for_mixed_vol}
\end{eqnarray}
where $x$ and $y$ are complex variables and $a,b,c,d,e$ are complex coefficients.
The CBB of this system is $4\cdot 3=12$, i.e., there can
be a maximum of $12$ isolated solutions for this system in $\mathbb{C}^{2}$.
The 2HomBB, using $\{x\}\times\{y\}$ is $6$.
Now, the supports of these equations are $S_{1}=\{(0,0),(1,0),(2,2)\}$
and $S_{2}=\{(0,0),(1,0),(0,1),(1,2)\}$ respectively. The Newton
polytope for $f_{1}(x,y)$ is $Q_{1}=\mbox{conv}(S_{1})=\{(0,0),(1,0),(1,1),(2,2)\}$
and for $f_{2}(x,y)$ it is $Q_{2}=\mbox{conv}(S_{2})=\{(0,0),(1,0),(0,1),(1,1),(1,2)\}$.

A Minkowski sum of any two convex sets is defined as
\begin{equation}
Q_{1}+Q_{2}=\{q_{1}+q_{2}\colon q_{1}\in Q_{1},q_{2}\in Q_{2}\}.
\end{equation}
The Minkowski sum of two Newton polytopes corresponds to multiplying the corresponding polynomials algebraically.
The $m$-dimensional volume of a simplex having vertices $v_{0},v_{1},\dots,v_{m}$, is
\begin{equation}
\mbox{Vol}_{m}(\mbox{conv}(v_{0},\dots,v_{m}))=\frac{1}{m!}|\det[v_{1}-v_{0},\dots,v_{m}-v_{0}]|.
\end{equation}
 From there on, one can show that the $m$-dimensional volume $\mbox{Vol}_{m}(\lambda_{1}Q_{1}+\dots+\lambda_{m}Q_{m})$,
where $0\ge\lambda_{i}\in\mathbb{R}$, is a homogeneous polynomial
of degree $m$ in variables $\lambda_{i}$. The mixed volume of convex
polytopes $Q_{1},\dots,Q_{m}$, denoted
$M(Q_{1},\dots,Q_{m})$, is defined as the coefficient of $\lambda_1\cdots\lambda_m$
in $\mbox{Vol}_{m}(\lambda_{1}Q_{1}+\dots+\lambda_{m}Q_{m})$.  It can be shown that
\begin{equation}
M(Q_{1},\dots,Q_{m})=\sum_{i=1}^{m}(-1)^{m-i}\mbox{ Vol}_{m}(\sum_{j\in\Omega_{i}^{m}}Q_{j}),
\end{equation}
where the inner sum is a Minkowski sum of polytopes and $\Omega_{i}^{m}$
are the combinations of $m$-objects (i.e., $m$-dimensional geometrical
objects made of $m$-simplices) taken $i$ at a time.  Moreover,
the mixed volume is always an integer for a system
of Laurent polynomials. For the case of two polynomials in two variables,
\begin{equation}
M(Q_{1},Q_{2})=\mbox{ Vol}_{2}(Q_{1}+Q_{2})-\mbox{ Vol}_{2}(Q_{1})-\mbox{ Vol}_{2}(Q_{2}).\label{eq:two_variable_mixvol}
\end{equation}

For the system in (\ref{example_system_for_mixed_vol}),
\begin{eqnarray}
\mbox{Vol}_{2}(Q_{1}) & = & 1,\nonumber \\
\mbox{Vol}_{2}(Q_{2}) & = & \mbox{ area of parallelogram made by \ensuremath{\{(0,0),(1,0),(0,1),(1,1)\}}}\nonumber \\
 & + & \mbox{ area of triangle made by \ensuremath{\{(1,1),(1,0),(1,2)\}}}\nonumber \\
 & = & 1+\frac{1}{2}=\frac{3}{2},\nonumber \\
\mbox{Vol}_{2}(Q_{1}+Q_{2}) & = & \frac{13}{2}.
\end{eqnarray}
Thus, the mixed volume for this system is $4$.
This is important because the
Bernstein theorem and Bernstein-Khovanskii-Kushnirenko (BKK) theorem~\cite{Bernstein75,Khovanski78,Kushnirenko76}
says that for generic coefficients, the number of isolated solutions in $(\mathbb{C}^{*})^{m}$
of a Laurent system is exactly equal to the mixed volume of this system
counting with multiplicity.  For any particular set of coefficients, this is an upper
bound. This result is very interesting since one can get a generically sharp bound
on the number of solutions in $(\mathbb{C}^{*})^{m}$ of a polynomial
system by knowing the exponent vectors of monomials.
\comment{This result originates
from toric varieties and related issues in Algebraic Geometry. We
do not intend to go into further details, rather we discuss its application
to the problems at hand.}
For the system in (\ref{example_system_for_mixed_vol}), there
can be a maximum of $4$ isolated solutions in $(\mathbb{C}^{*})^{2}$, which is in contrast
to the CBB and 2HomBB bound of $12$ and $6$, respectively, in $\mathbb{C}^2$.

To yield the BKK count for $\mathbb{C}^{m}$, one considers an extension of
the mixed volume, called the \textit{stable mixed volume}
\footnote{Because of the highly technical nature of the stable mixed volume,
the bound in $\IC^{m}$ which has been commonly used and implemented
in the community is the bound given in \cite{Li96thebkk}. This
bound is quite easy to state: Add a constant term to polynomials in
the system which do not have constant term, and the mixed volume of
the resulting augmented system serves as a bound in $\IC^{m}$ for the
original system. The {}``stable mixed volume\textquotedblright{}
is a little bit more general then this {}``augmented mixed volume\textquotedblright{}.
We would like to thank T.Y. Li for clarifying this point.},
which ensures that we have all necessary solutions in
$\mathbb{C}^{m}$ \cite{Li96thebkk,Mau:94,MauW:96}.
A discussion on the stable mixed volume is beyond the scope of this article, although
its calculation is similar to that of the mixed volume with some formal
complications. However, it should be noted that a highly sophisticated
implementation of an algorithm to calculate the mixed volume of a
given system is MixedVol \cite{GLW:05} which is transplanted in {\sf PHCpack}
and MixedVol-2.0 which is transplanted in {\sf HOM4PS2}.

Now, after calculating the stable mixed volume of the original system
$P(x)=0$, we require a start system, $Q(x)=0$,
such that $Q(x)$ has the same stable mixed volume, where of course
solutions of $Q(x)=0$ should be known or can be obtained easily.
This homotopy method is called Polyhedral Homotopy.

\newpage

\bibliographystyle{unsrt}
\bibliography{bibliography_NPHC_NAG}

\end{document}